\documentclass[12pt]{article}
\usepackage{fancyhdr}
\usepackage{titling} % Optional: for additional control of the title layout
\usepackage{lipsum}  % For placeholder text (can be removed)

\fancypagestyle{titlepageheader}{
  \fancyhf{} % Clear all headers and footers
  \fancyhead[C]{Yau Science Award 2024 Computer Science} % Center-aligned header text
}

\usepackage[margin=1in]{geometry}
\usepackage{graphicx}
\usepackage{subfigure}
\usepackage{placeins}
\usepackage{subfigure} 
\usepackage{braket}
\usepackage{placeins}
\usepackage[english]{babel}
\usepackage[utf8]{inputenc}
\usepackage{subcaption}
\usepackage{amsmath}
\usepackage{amssymb}
\usepackage{amsfonts}
\usepackage{amsthm}
\usepackage{mathdots}
\usepackage{mathrsfs}
\usepackage{setspace}
\usepackage[all]{xy}
\usepackage{float}
\usepackage{color}
\usepackage{cite}
\usepackage{url}
\usepackage{indent first}
\usepackage[labelfont=bf,labelsep=period,justification=raggedright]{caption}
\usepackage[english]{babel}
\usepackage[utf8]{inputenc}
\usepackage{hyperref}
\usepackage[colorinlistoftodos]{todonotes}
\usepackage{tikz}
\usepackage{tikz-3dplot}
\usetikzlibrary{calc}
\usetikzlibrary{knots}
\usetikzlibrary{spath3, intersections, hobby}
\usetikzlibrary{arrows, automata}
\usetikzlibrary{arrows.meta}
\usetikzlibrary{shapes.multipart}
\usepackage{pgfplots}
\usepackage{multicol}
\PassOptionsToPackage{dvipsnames,svgnames}{xcolor}
\usepackage{textcomp}
\usepackage{bbm}
\usepackage{array}
\newcolumntype{C}{>{$}c<{$}}
\usetikzlibrary{3d}
\tdplotsetmaincoords{80}{110}
\usepackage{multicol}

\makeatletter
\DeclareFontFamily{OMX}{MnSymbolE}{}
\DeclareSymbolFont{MnLargeSymbols}{OMX}{MnSymbolE}{m}{n}
\SetSymbolFont{MnLargeSymbols}{bold}{OMX}{MnSymbolE}{b}{n}
\DeclareFontShape{OMX}{MnSymbolE}{m}{n}{
    <-6>  MnSymbolE5
   <6-7>  MnSymbolE6
   <7-8>  MnSymbolE7
   <8-9>  MnSymbolE8
   <9-10> MnSymbolE9
  <10-12> MnSymbolE10
  <12->   MnSymbolE12
}{}
\DeclareFontShape{OMX}{MnSymbolE}{b}{n}{
    <-6>  MnSymbolE-Bold5
   <6-7>  MnSymbolE-Bold6
   <7-8>  MnSymbolE-Bold7
   <8-9>  MnSymbolE-Bold8
   <9-10> MnSymbolE-Bold9
  <10-12> MnSymbolE-Bold10
  <12->   MnSymbolE-Bold12
}{}

\let\llangle\@undefined
\let\rrangle\@undefined
\DeclareMathDelimiter{\llangle}{\mathopen}%
                     {MnLargeSymbols}{'164}{MnLargeSymbols}{'164}
\DeclareMathDelimiter{\rrangle}{\mathclose}%
                     {MnLargeSymbols}{'171}{MnLargeSymbols}{'171}
\makeatother

\makeatletter
\tikzoption{canvas is xy plane at z}[]{%
  \def\tikz@plane@origin{\pgfpointxyz{0}{0}{#1}}%
  \def\tikz@plane@x{\pgfpointxyz{1}{0}{#1}}%
  \def\tikz@plane@y{\pgfpointxyz{0}{1}{#1}}%
  \tikz@canvas@is@plane
}
\makeatother

\tikzset{surface/.style={draw=black, left color=orange,right color=orange,middle
color=orange!60!#1, fill opacity=1},surface/.default=white}

\tikzset{>=latex}

\pgfplotsset{compat=1.17}

\makeatletter
\newcommand{\tpitchfork}{%
  \vbox{
    \baselineskip\z@skip
    \lineskip-.52ex
    \lineskiplimit\maxdimen
    \m@th
    \ialign{##\crcr\hidewidth\smash{$-$}\hidewidth\crcr$\pitchfork$\crcr}
  }%
}
\makeatother

\def\multiset#1#2{\ensuremath{\left(\kern-.3em\left(\genfrac{}{}{0pt}{}{#1}{#2}\right)\kern-.3em\right)}}

\makeatletter

\newif\ifpgfcirclecrosssplitcustomfill
\tikzset{%
  circle cross split part fill/.code=\def\pgf@lib@sh@ccs@list@fill{#1}\pgfcirclecrosssplitcustomfilltrue,%
  circle cross split uses custom fill/.is if=pgfcirclecrosssplitcustomfill}
\pgfdeclareshape{circle cross split}{%
  \nodeparts{text,two,three,four}%
  \savedanchor\centerpoint{%
    \pgfmathsetlength\pgf@xa{\pgfkeysvalueof{/pgf/inner xsep}}%
    \pgfmathsetlength\pgf@ya{\pgfkeysvalueof{/pgf/inner ysep}}%
    \pgf@x\wd\pgfnodeparttextbox
    \pgf@yb\dp\pgfnodeparttextbox
    \pgf@yc\dp\pgfnodeparttwobox
    \ifdim\pgf@yb>\pgf@yc
      \pgf@yc\pgf@yb
    \fi
    \advance\pgf@y-\pgf@yc
    \advance\pgf@x\pgf@xa
    \advance\pgf@y-\pgf@ya
    \advance\pgf@x.5\pgflinewidth
    \advance\pgf@y-.5\pgflinewidth
  }%
  \savedanchor\twoanchor{%
    \pgfmathsetlength\pgf@xa{\pgfkeysvalueof{/pgf/inner xsep}}%
    \pgfmathsetlength\pgf@ya{\pgfkeysvalueof{/pgf/inner ysep}}%
    \advance\pgf@x.5\pgflinewidth
    \advance\pgf@x\pgf@xa
    \advance\pgf@y.5\pgflinewidth
    \advance\pgf@y\pgf@ya
    \pgf@yb\dp\pgfnodeparttextbox
    \pgf@yc\dp\pgfnodeparttwobox
    \ifdim\pgf@yb>\pgf@yc
      \pgf@yc\pgf@yb
    \fi
    \advance\pgf@y\pgf@yc
  }%
  \savedanchor\threeanchor{%
    \pgfmathsetlength\pgf@ya{\pgfkeysvalueof{/pgf/inner ysep}}%
    \pgf@x\wd\pgfnodeparttextbox
    \pgf@yb\dp\pgfnodeparttextbox
    \pgf@yc\dp\pgfnodeparttwobox
    \ifdim\pgf@yb>\pgf@yc
      \pgf@yc\pgf@yb
    \fi
    \advance\pgf@y-\pgf@yc
    \advance\pgf@y-2\pgf@ya
    \advance\pgf@y-\pgflinewidth
    \pgf@yb\ht\pgfnodepartthreebox
    \pgf@yc\ht\pgfnodepartfourbox
    \ifdim\pgf@yb>\pgf@yc
      \pgf@yc\pgf@yb
    \fi
    \advance\pgf@y-\pgf@yc
    \advance\pgf@x-\wd\pgfnodepartthreebox
  }%
  \savedanchor\fouranchor{%
    \pgfmathsetlength\pgf@xa{\pgfkeysvalueof{/pgf/inner xsep}}%
    \advance\pgf@x\wd\pgfnodepartthreebox
    \advance\pgf@x2\pgf@xa
    \advance\pgf@x\pgflinewidth
  }%
  \saveddimen\radius{%
    \pgf@y\ht\pgfnodeparttextbox
    \pgf@yb\ht\pgfnodeparttwobox
    \ifdim\pgf@yb>\pgf@y
      \pgf@y\pgf@yb
    \fi
    \pgf@yc\dp\pgfnodeparttextbox
    \pgf@yb\dp\pgfnodeparttwobox
    \ifdim\pgf@yc>\pgf@yb
      \advance\pgf@y\pgf@yc
    \else
      \advance\pgf@y\pgf@yb
    \fi
    \pgf@yb\ht\pgfnodepartthreebox
    \ifdim\pgf@yb<\ht\pgfnodepartfourbox
      \pgf@yb\ht\pgfnodepartfourbox
    \fi
    \pgf@yc\dp\pgfnodepartthreebox
    \ifdim\pgf@yc<\dp\pgfnodepartfourbox
      \advance\pgf@yb\dp\pgfnodepartfourbox
    \else
      \advance\pgf@yb\pgf@yc
    \fi
    \ifdim\pgf@yc>\pgf@y
      \pgf@y\pgf@yc
    \fi
    \pgfmathsetlength\pgf@ya{\pgfkeysvalueof{/pgf/inner ysep}}%
    \advance\pgf@y2\pgf@ya
    \pgf@x\wd\pgfnodeparttextbox
    \pgf@xa\wd\pgfnodepartthreebox
    \pgf@xb\wd\pgfnodeparttwobox
    \pgf@xc\wd\pgfnodepartfourbox
    \ifdim\pgf@xa>\pgf@x
      \pgf@x\pgf@xa
    \fi
    \ifdim\pgf@xb>\pgf@x
      \pgf@x\pgf@xb
    \fi
    \ifdim\pgf@xc>\pgf@x
      \pgf@x\pgf@xc
    \fi
    \pgfmathsetlength\pgf@xa{\pgfkeysvalueof{/pgf/inner xsep}}%
    \advance\pgf@x2\pgf@xa
    \ifdim\pgf@y>\pgf@x
      \pgf@x\pgf@y
    \fi
    \advance\pgf@x.5\pgflinewidth
    \pgfmathsetlength{\pgf@xb}{\pgfkeysvalueof{/pgf/minimum width}}%
    \pgfmathsetlength{\pgf@yb}{\pgfkeysvalueof{/pgf/minimum height}}%
    \ifdim\pgf@x<.5\pgf@xb
        \pgf@x=.5\pgf@xb
    \fi
    \ifdim\pgf@x<.5\pgf@yb
        \pgf@x=.5\pgf@yb
    \fi
    \pgfmathsetlength{\pgf@xb}{\pgfkeysvalueof{/pgf/outer xsep}}%
    \pgfmathsetlength{\pgf@yb}{\pgfkeysvalueof{/pgf/outer ysep}}%
    \ifdim\pgf@xb<\pgf@yb
      \advance\pgf@x\pgf@yb
    \else
      \advance\pgf@x\pgf@xb
    \fi
  }%
  \inheritanchorborder[from=circle]%
  \inheritanchor[from=circle]{north}%
  \inheritanchor[from=circle]{north west}%
  \inheritanchor[from=circle]{north east}%
  \inheritanchor[from=circle]{center}%
  \inheritanchor[from=circle]{west}%
  \inheritanchor[from=circle]{east}%
  \inheritanchor[from=circle]{mid}%
  \inheritanchor[from=circle]{mid west}%
  \inheritanchor[from=circle]{mid east}%
  \inheritanchor[from=circle]{base}%
  \inheritanchor[from=circle]{base west}%
  \inheritanchor[from=circle]{base east}%
  \inheritanchor[from=circle]{south}%
  \inheritanchor[from=circle]{south west}%
  \inheritanchor[from=circle]{south east}%
  \anchor{two}{\twoanchor}%
  \anchor{three}{\threeanchor}%
  \anchor{four}{\fouranchor}%
  \inheritbackgroundpath[from=circle]
  \beforebackgroundpath{%
    \pgfutil@tempdima=\radius
    \pgfmathsetlength{\pgf@xb}{\pgfkeysvalueof{/pgf/outer xsep}}%  
    \pgfmathsetlength{\pgf@yb}{\pgfkeysvalueof{/pgf/outer ysep}}%  
    \ifdim\pgf@xb<\pgf@yb
      \advance\pgfutil@tempdima by-\pgf@yb
    \else
      \advance\pgfutil@tempdima by-\pgf@xb
    \fi
    \advance\pgfutil@tempdima by-.5\pgflinewidth%
    \pgfsetshortenstart{0pt}%
    \pgfsetshortenend{0pt}%
    \pgfsetarrows{-}%  
    \pgfpathmoveto{\pgfpointadd{\centerpoint}{\pgfqpoint{-\pgfutil@tempdima}{0pt}}}%
    \pgfpathlineto{\pgfpointadd{\centerpoint}{\pgfqpoint{\pgfutil@tempdima}{0pt}}}%
    \pgfpathmoveto{\pgfpointadd{\centerpoint}{\pgfqpoint{0pt}{-\pgfutil@tempdima}}}%
    \pgfpathlineto{\pgfpointadd{\centerpoint}{\pgfqpoint{0pt}{\pgfutil@tempdima}}}%
    \pgfusepathqstroke
  }%
  \behindbackgroundpath{%
    \pgfutil@tempdima=\radius
    \pgfmathsetlength{\pgf@xb}{\pgfkeysvalueof{/pgf/outer xsep}}%  
    \pgfmathsetlength{\pgf@yb}{\pgfkeysvalueof{/pgf/outer ysep}}%  
    \ifdim\pgf@xb<\pgf@yb
      \advance\pgfutil@tempdima by-\pgf@yb
    \else
      \advance\pgfutil@tempdima by-\pgf@xb
    \fi
    \advance\pgfutil@tempdima by-.5\pgflinewidth%
    \ifpgfcirclecrosssplitcustomfill%
      \pgf@lib@sh@rs@process@list{\pgf@lib@sh@ccs@list@fill}{4}%
      {%
        \pgfmathloop
           \ifnum\pgfmathcounter>4%
           \else%
             \pgf@lib@sh@getalpha\pgf@lib@sh@rs@number{\pgfmathcounter}%
              \edef\pgf@tempa{\csname pgf@lib@sh@rs@\pgf@lib@sh@rs@number @item\endcsname}%
              \ifx\pgf@tempa\pgf@lib@sh@rs@nonetext\else
                \pgfsetfillcolor{\pgf@tempa}%
                \pgf@lib@sh@ccs@angles{\pgfmathcounter}%
                \pgfpathmoveto{\centerpoint}%
                \pgfpathlineto{\pgfpointadd{\centerpoint}{\pgfqpointpolar{\pgf@lib@sh@ccs@angle}{\pgfutil@tempdima}}}%
                \pgfpatharc{\pgf@lib@sh@ccs@angle}{\pgf@lib@sh@ccs@angle@}{\pgfutil@tempdima}%
                \pgfpathclose
                \pgfusepathqfill
              \fi
        \repeatpgfmathloop
      }%
    \fi%
  }%
}
\def\pgf@lib@sh@ccs@angles#1{%
  \ifcase#1\or\def\pgf@lib@sh@ccs@angle{90}%
           \or\def\pgf@lib@sh@ccs@angle{0}%
           \or\def\pgf@lib@sh@ccs@angle{180}%
           \else\def\pgf@lib@sh@ccs@angle{270}%
  \fi
  \edef\pgf@lib@sh@ccs@angle@{\number\numexpr\pgf@lib@sh@ccs@angle+90\relax}%
}
\makeatother

\theoremstyle{plain}
\newtheorem{thm}{Theorem}

\theoremstyle{definition}
\newtheorem{defn}[thm]{Definition}

\theoremstyle{remark}

\numberwithin{equation}{section}
\numberwithin{thm}{section}

\author{Alejandro Antonio Mayorga, Alexander Yuan}

\date{}
\title{CTRQNets \& LQNets: Continuous Time Recurrent and Liquid Quantum Neural Networks}
\begin{document}
\raggedbottom
\begin{figure}[H]
    \centering
    \includegraphics[width=\textwidth]{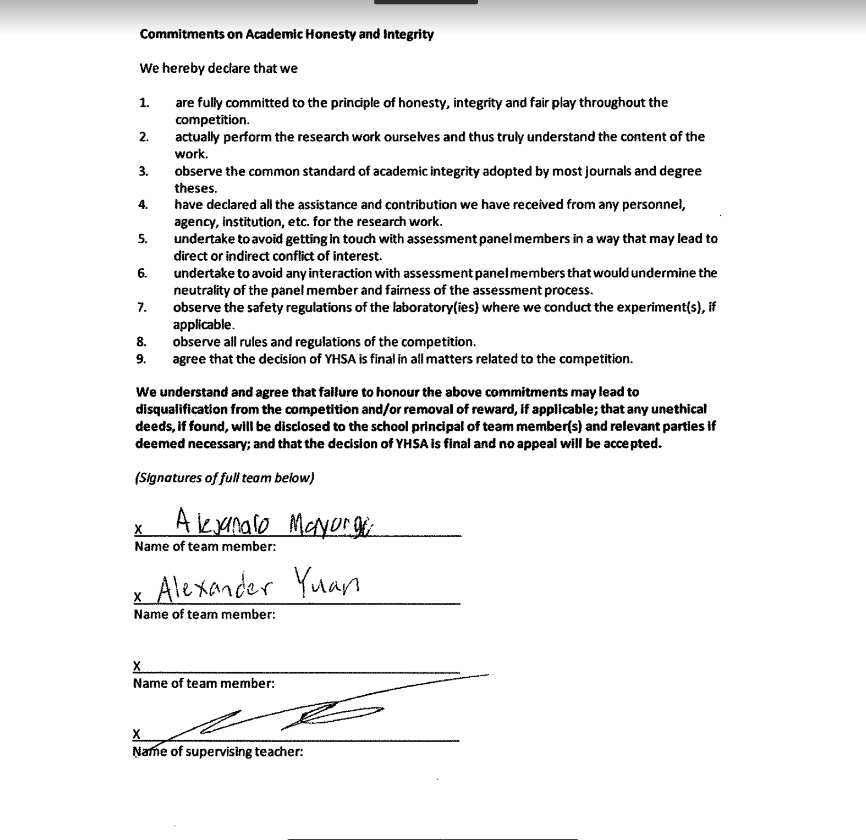}  % Replace with your figure filename
    \caption*{S.T Yau Science Award Computer Science 2024}  % Creates an unnumbered caption
\end{figure}

\maketitle
\begin{center}

    \textbf{Mentors:} Tyler Wooldridge, Andrew Yuan
    
     \textbf{School:} Danbury Math Academy Danbury, Connecticut, United States of America

    \textbf{Advisor:} Dr. Xiaodi Wang
    
\end{center}

\newpage
\pagestyle{plain} % Default style for remaining pages
\section*{Abstract}
Neural networks have continued to gain prevalence in the modern era for their ability to model complex data through pattern recognition and behavior remodeling. However, the static construction of traditional neural networks inhibits dynamic intelligence. This makes them inflexible to temporal changes in data and unfit to capture complex dependencies. With the advent of quantum technology, there has been significant progress in creating quantum algorithms. In recent years, researchers have developed quantum neural networks that leverage the capabilities of qubits to outperform classical networks. However, their current formulation exhibits a static construction limiting the system's dynamic intelligence. To address these weaknesses, we develop a Liquid Quantum Neural Network (LQNet) and a Continuous Time Recurrent Quantum Neural Network (CTRQNet). Both models demonstrate a significant improvement in accuracy compared to existing quantum neural networks (QNNs), achieving accuracy increases as high as 40\% on CIFAR 10 through binary classification. We propose LQNets and CTRQNets might shine a light on quantum machine learning's black box.

%However, the static construction of traditional neural networks reduces efficiency in capturing complex dependencies and causes such models to be inflexible to temporal changes in data. High computational costs and decreased adaptability limit the effectiveness and implementation of machine learning technology. 

 \emph{Keywords: Liquid Networks, Quantum Computing, QNNs, CTRNets, LQNet, CTRQNet}

\section*{Highlights}
In this paper, we create new classes of models, namely the LQNet and CTRQNet, that take advantage of the benefits of quantum computing to analyze data more effectively. These new models are more dynamic and can apply patterns learned from data to broader datasets. As a result, these models can find meaningful patterns in datasets and learn complex relationships between data more effectively. Compared to a QNN, LQNets and CTRQNets are able to learn faster, reaching an optimal state in less than half of the iterations typical of QNNs. For some datasets, LQNets and CTRQNets can find patterns that QNNs cannot, allowing for more general usage of AI.

\newpage
\section*{Contributions of Authors}
Alejandro Mayorga developed the theoretical framework for the LQNet and CTRQNet. He developed the main code for these models and performed the baseline testing.

Alex Yuan implemented and optimized the base QNN model. He conducted the analysis and comparison of the network results.
\section*{Acknowledgements}
The authors wish to give special thanks to advisor Dr. Wang for his extensive help on the project and for guiding our research as well as providing resources. This project would not have been possible without him, and he serves as an inspiration to all of us.

The authors would also like to thank mentors Tyler Wooldridge and Andrew Yuan for their extensive contributions in formulating the models. Their hard work, enthusiasm, and outstanding insights deeply motivated us, and we are incredibly grateful for their support in this endeavor.

\newpage

\begin{small}
\begin{spacing}{0}
\subsection*{Commitments on Academic Honesty and Integrity}
We hereby declare that we
\begin{enumerate}
  \setlength{\itemsep}{0pt}
  \item are fully committed to the principle of honesty, integrity and fair play throughout the competition.
  \item actually perform the research work ourselves and thus truly understand the content of the work.
  \item observe the common standard of academic integrity adopted by most journals and degree theses.
  \item have declared all the assistance and contribution we have received from any personnel, agency, institution, etc. for the research work.
  \item undertake to avoid getting in touch with assessment panel members in a way that may lead to direct or indirect conflict of interest.
  \item undertake to avoid any interaction with assessment panel members that would undermine the neutrality of the panel member and fairness of the assessment process.
  \item observe the safety regulations of the laboratory(ies) where we conduct the experiment(s), if applicable.
  \item observe all rules and regulations of the competition.
  \item agree that the decision of YHSA is final in all matters related to the competition.
\end{enumerate}
\end{spacing}
\vspace{0.5cm}
\begin{spacing}{0.3}
\textbf{We understand and agree that failure to honour the above commitments may lead to disqualification from the competition and/or removal of reward, if applicable; that any unethical deeds, if found, will be disclosed to the school principal of team member(s) and relevant parties if deemed necessary; and that the decision of YHSA is final and no appeal will be accepted.} (Signatures of full team below) \\
\end{spacing}

X\underline{\hspace{8cm}} \\
Name of team member: \\
X\underline{\hspace{8cm}} \\
Name of team member:\\
X\underline{\hspace{8cm}} \\
Name of team member:\\
X\underline{\hspace{8cm}} \\
Name of supervising teacher:\\
X\underline{\hspace{8cm}} \\
Name of supervising teacher:\\
X\underline{\hspace{8cm}} \\
Name of supervising teacher:\\
\end{small}
\newpage
\tableofcontents
\newpage
\section{Introduction}

\subsection{Prior Works}
Machine learning is the development of intelligence through mathematical tools. This field has garnered increasing media attention over the past two decades due to the scalability of machine learning methods \cite{javapoint}. While large language models such as ChatGPT \cite{chatgpt1,chatgpt2,facultyfocus_chatgpt} are the most publicized applications of machine learning, they do not encompass the entire field. Moreover, machine learning itself is merely a sub-field of the more well-known field of artificial intelligence, which is more broadly concerned with developing ``intelligence" through either biological processes (synthetic biological intelligence)\cite{bio1,bio2,bio3,stanford_hai_biological_intelligence} or mathematical and computer processes (machine learning). While both methods show promise, the rich history of mathematics provides a deeper understanding of how intelligence may be simulated. Specifically, neural networks \cite{mitnn,ibmnn} are notable for their ability to learn abstract patterns even in large datasets, a property that does not extend to other methods, such as support vector machines (SVMs) \cite{svm1}.

Deep learning is the development of algorithms that are closely related to neural networks. Examples of this are seen in many real-world applications [See \cite{chatgpt1,chatgpt2,ibm_cnn,NODES}]. These models only depend on one set of parameters at any given time, which is key to their scalability. Additionally, the sheer size of the parameters that characterize these models makes understanding what these networks truly ``know" an enigmatic task. Thus, controlling and assessing their capabilities becomes difficult \cite{ibm_explainable_ai,vox_unexplainable_chatgpt}. 

In 1993, a continuous-time recurrent neural network (CTRNN) developed by \cite{1993} proved that ``that any finite time trajectory of a given n-dimensional dynamical system can be approximately realized by the internal state of the output units of a continuous time recurrent neural network with $n$ output units, some hidden units, and an appropriate initial condition." CTRNNs have been applied to evolutionary robotics, where they have been used to address vision, cooperation, and minimal cognitive behavior \cite{wiki}.

However, the recent development of ``liquid models"\cite{LNNS} has made this task much more feasible. The backbone behind the construction of liquid models lies in the fact that building a model with dynamic intelligence and task awareness will vastly simplify controlling and assessing the model's knowledge. For example, a car was successfully operated using only 19 neurons of a liquid neural network, whereas traditional LSTMs would require millions of neurons to perform this same task \cite{LSTM}. Liquid models have a much closer representation of what is understood to be intelligence, as these models are dynamic and task-aware even past the training process.

\subsection{Our new ideas}
Liquid networks\cite{LNNS} and continuous time recurrent neural networks (CTRNets)\cite{1993} have shown great promise in displaying dynamic intelligence, a feature that classical networks lack. While these models have shown great promise in current literature, no work has been done to investigate a continuous hidden state in quantum neural networks (QNNs). The current development of QNNs has shown to be superior, theoretically, to classical networks. However, these models still admit a rigid hidden state leading to static intelligence.

To address these issues, we fuse these continuous networks and quantum neural networks, introducing Liquid Quantum Networks (LQNets) and Continuous-Time Recurrent Quantum Neural Networks (CTRQNets).  Firstly, we define a quantum residual block from \cite{QRESNET} to construct a residual block to mitigate the vanishing gradient. Inspired by \cite{NODES}, we derive a new model for quantum neural ordinary differential equations, which we then use to formulate an LQNet and CTRQNet. These models surpass quantum neural networks by capturing temporal dependencies and exhibiting dynamic intelligence through an ever-changing set of differential equations governing there hidden state. On CIFAR-10, our models achieved a 40\% increase in accuracy compared to QNNs.
\section{Preliminaries}

\subsection{Multi-layer Neural Networks}
%In the current literature, there is a lack of a rigorous definition of a neural network. Throughout this paper, we develop and utilize neural networks as dynamical systems similar to \cite{LNNS,NODES,Hopfield}. Within the field of dynamical systems, a substantial amount of work has been done, so we leverage this knowledge and apply it to neural networks. In order to use methods from the field of dynamical systems, it is necessary to formulate a well-defined and rigorous definition for a neural network. While there is a wide range of neural networks \cite{LSTM,GRU,Resnet,CNNS} most of these model architectures are derived from the multi-layer perceptron introduced by \cite{NNS}. It's of utmost importance to define a perceptron as it will enable the exploration of neural networks as dynamical systems. Any definition which will be of any use should encapsulate the following elements of a neural network:
%\begin{enumerate}
    %\item Forward pass of input through hidden layers.
    %\item Calculation of loss throughout the network.
    %\item Backpropagation of the loss throughout the network by calculating the gradient of the loss function with respect to the model's parameters.
    %\item The update of the model's parameters.
%\end{enumerate}
We define a neural network as follows:
\begin{defn}
   If $\mathcal{F}(x)$ be a composition from the set $\{f_n\}$ where $\{f_n\}$ denotes a finite set of differentiable functions on $\mathbb{R}$, then $\mathcal{F}$ is a neural network.
\end{defn}
However, this definition only considers static neural networks, not cases where hidden states are solutions to differential equations. Mathematically defining networks with these implicit hidden states is challenging due to the various methods of forward and backward propagation used in different architectures \cite{Adjoint,Bogacki–Shampine,LNNS,NODES}. For example, \cite{LNNS} uses a backpropagation through a time training algorithm, whereas \cite{NODES} uses an adjoint method with the ladder being the continuous counterpart of the former. These formulations are mathematically different and, therefore, cannot be encompassed under the same definition. We omit a definition for these types of networks but define \emph{dynamical, continuous, liquid hidden states} to be hidden states where the corresponding transformation is the solution to a differential equation.\\
The current literature has significant room to explore continuous dynamics in quantum neural networks as their static construction resembles that of their classical counterparts. This static construction is sub-optimal for real-world applications where data may be irregularly spaced, leading to quantum networks' inability to learn from this type of data in their current formulation. By replacing the static hidden states in quantum networks with a dynamic liquid structure, these newly formulated quantum models, which we call CTRQNets and LQNets, are able to adapt to various types of data that quantum neural networks struggle to adapt to. These models exhibit dynamical intelligence and are task-aware even after the training process is complete, so the knowledge of these models can be easily accessed.

\subsection{Residual Learning}

A common issue seen in deep neural networks is the degradation of accuracy as the number of layers increases. This issue is not due to overfitting, as deeper networks report lower training accuracy than shallower ones on certain tasks\cite{Resnet}. To address this, \cite{Resnet} proposed residual networks where layers learn the residual between the optimal hidden state and the input rather than directly learning the optimal hidden state. As these layers are designed to learn the residual mapping, there is no degradation in accuracy with additional layers. If a layer fails to learn a meaningful pattern from the data, it defaults to an identity mapping, ensuring that it does not negatively impact the network's performance.

Residual networks operate under the assumption that stacked layers can asymptotically approximate complex functions. While this is a strong claim, there is evidence supporting this idea \cite{1993,LNNS}. Let $\mathcal{H}(x)$ denote the optimal series of transformation, and let $\mathcal{F}(x) := \mathcal{H}(x) - x$ be the residual function corresponding to the optimal series of transformations. Additionally, let $f_t(x)$ denote the transformations from the weights and biases at optimization step $t$. Define $h_t(x) = f_t(x)+x$ where $h_t(x)$ is the output of the block. This new formulation for the output of the block enables $h_t(x)\to \mathcal{H}(x)$ and therefore $f(x)\to \mathcal{F}(x)$ Therefore, the optimal series of transformations from the weight and biases corresponds to the residual $\mathcal{F}(x) := \mathcal{H}(x) - x$

Key advantages of residual networks:
\begin{enumerate}
    \item Easier optimization compared to vanilla networks.
    \item Ability to add more layers without accuracy degradation.
\end{enumerate}

\begin{figure}
    \centering
    \includegraphics{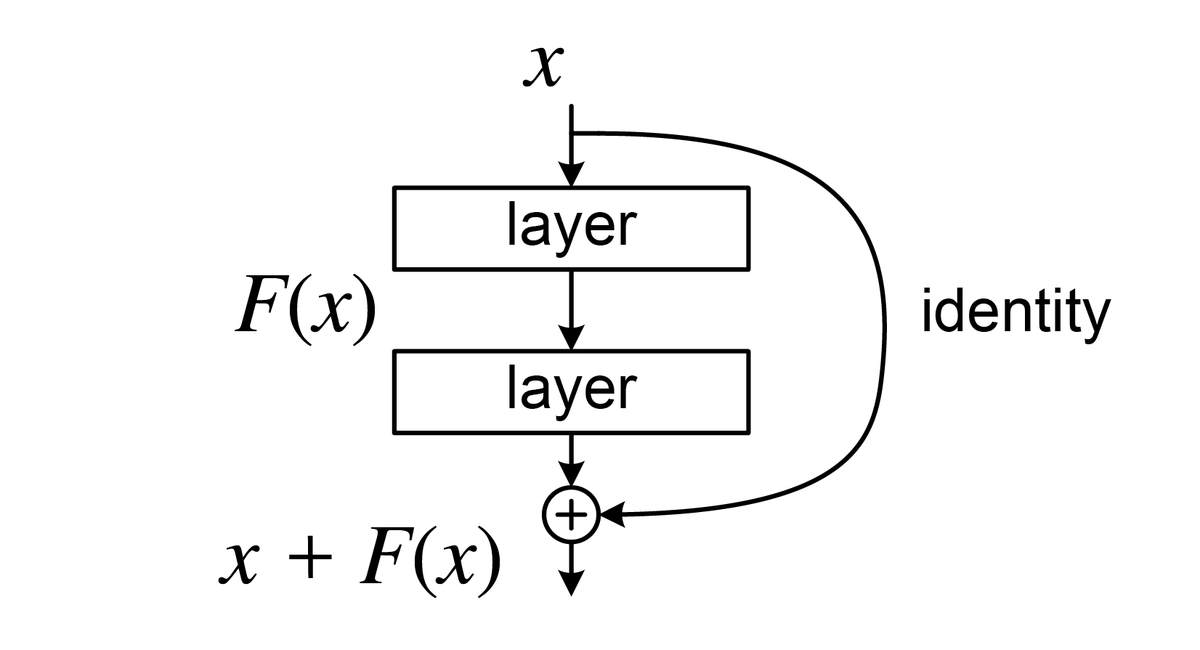}
    \caption{\label{fig:ResidualBlock}The Residual Block}
\end{figure}

\subsection{Neural Ordinary Differential Equations (NODES)}

\subsubsection{Neural Ordinary Differential Equations}
Consider a residual block of the form $$h_{t+1} = h_t+f(\theta_t,h_t,t)$$
where $h_t$ denotes the $t$-th residual block within the layer and $\theta_t$ denotes the model parameters corresponding to the $t$-th residual block within the network. Residual networks are discrete function approximators as they lack a continuous structure. Discrete neural networks \emph{i.e. residual networks} have 2 main disadvantages:
\begin{enumerate}
    \item They lack the adaptability to process non-rigid data (time-series data with non-constant time spacing).
    \item They lack dynamic intelligence and cannot self-adjust to unfamiliar patterns due to the fact that the hyperparameters are fixed.

\end{enumerate}
In 2018, the authors of \cite{NODES} formulated a solution to this problem, which showed great promise for future work. Euler's method for solving ordinary differential equations (ODES) states that  given $$\frac{dy}{dt} = f(t,y)$$
subject to the initial condition $y(t_0) = y_0$ an approximation of the solution is given by $$y_{n+1} = y_n+hf(t_n,y_n)$$ where $h$ is the step size and  $(t_n,y_n)$ lies on the approximation curve. If we consider $h=1$, the above equation resembles that of a residual block. In theory, by adding more layers, we allow for a continuous neural network where there exist arbitrary layers, \emph{i.e., the network is defined at arbitrary layers.} Such a network resembles a continuous function $f$ such that
\begin{equation}
    \frac{d\textbf{z}(t)}{dt} = f(\textbf{z}(t),t,\theta)
\end{equation}
If $\textbf{z}(0)$ denotes the input layer of the network and $\textbf{z}(T)$ denotes the output layer of the network, we have $$\int_{0}^{T}f(\textbf{z}(t),t,\theta)dt + \textbf{z}(0) = \textbf{z}(T).$$
In order to compute the forward propagation of the network, we solve Equation 2.1. The hidden state is not predefined but rather dependent on the specific sample. Thus, the network is task-aware, inhibited by the static construction of residual networks.

\subsubsection{Numerical Methods}
There exist a variety of methods to perform gradient calculation, but the authors of \cite{NODES} used a method formulated in \cite{Adjoint}, which makes the calculation of these values explicit and preserves the continuous structure of Neural ODEs. While the adjoint rapidly performs gradient calculation, many intermediate optimization steps are taken due to the flexibility of ODE Solvers\cite{Bogacki–Shampine,Adjoint}, slowing the training process. In most cases\cite{LNNS,LSTM,GRU}, backpropagation through time is used, which is faster and more accurate in practice.

When minimizing loss function \(\mathcal{L}\) at \(z(T)\), \(\mathcal{L}\) depends not only on \(z(T)\) but on all \(z(t)\) for \(t \in [0, T]\). Neural ODEs have continuous depth, making direct backpropagation through these layers impossible. Instead, we consider partitioning the interval \([0, T]\) and updating the function throughout these intervals. Adaptive ODE solvers, such as Runge-Kutta and Bogacki–Shampine methods \cite{Runge, Bogacki–Shampine}, adjust intervals based on the smoothness of the solution. Larger steps are taken in smooth regions where the accuracy in the region is not critical towards the final solution, while smaller steps are used in volatile regions where accuracy is critical \cite{Press, Bogacki–Shampine, Runge}.

However, these ODE solvers cannot calculate the gradients directly, and using an ODE Solver to calculate the gradient would be very inefficient as it would have to solve the implicit differential equation:$$\mathcal{L}(z(t_1)) = \mathcal{L}({z(t_0) +\int_{t_0}^{t_1}f(\theta,z(t),t) dt)})$$
Instead, the adjoint method makes gradient computation much more explicit and allows adaptive ODE solvers to perform effectively. In Appendix A, we provide an overview of the adjoint method.

\subsection{Time Continuous Recurrent Neural Networks}
Continuous Time Recurrent Neural Networks (CTRNNs) had its first formulation \cite{1993} demonstrating the ability to approximate any continuous curve to any degree of accuracy. Like Recurrent Neural Networks, the hidden states of CTRNNs depend on time. A CTRNN is defined by the following differential equation:

\begin{equation}
    \frac{dx}{dt} = -\frac{1}{\tau}x(t) + W \sigma(x(t))
\end{equation}

To perform the forward pass of the network, the differential equation above is solved in the interval \( t = t_0 \) to \( t = t_1 \), subject to the initial condition of our input. In general, letting \( \mathbf{I}(t) \) denote the input to the CTRNN at time \( t \), we solve Eq. (2.1) in the interval $[t_{0}, t_{1}]$ subject to the initial condition $x(t_{0}) = \mathbf{I}(t_{0})$ for which $\sigma$ denotes a nonlinear activation function that is continuous and bounded (e.g., sigmoid). CTRNNs are more expressive than traditional networks due to the system of differential equations underlying their hidden state. These differential equations can be optimized through various numerical differential equation solvers, allowing CTRNNs to trade speed for accuracy in regions where it is necessary.

The following theorems were proven in \cite{1993}, creating the catalyst for more advanced research of differential equations and structure machine learning model structures. Theorems 2.2 and 2.3 below establish that the internal state of the output units of a CTRNN can approximate a finite time trajectory of a given dynamical system to an arbitrary precision:
\begin{thm}
Let $\sigma$ be a strictly increasing $C^1$-sigmoid function such that $\sigma(\mathbb{R}) = (0, 1)$, and let $f: I = [0, T] \rightarrow (0, 1)^n$ be a continuous curve, where $0 < T < \infty$. Then, for an arbitrary $\epsilon > 0$, there exists an integer $N$ and a recurrent neural network with $n$ output units and $N$ hidden units such that
\[
\max_{t \in I} \|f(t) - y(t)\| < \epsilon,
\]
where $y(t) = (\sigma(y_1(t)), \ldots, \sigma(y_n(t)))$ is the output of the recurrent network with the sigmoid output function $\sigma$.
\end{thm}
\begin{thm}
Let $f: I = [0, T] \rightarrow \mathbb{R}^n$ be a continuous curve, where $0 < T < \infty$. Then, for an arbitrary $\epsilon > 0$, there exist an integer $N$ and a recurrent network with $n$ output units and $N$ hidden units such that
\[
\max_{t \in I} \|f(t) - u(t)\| < \epsilon,
\]
where $u(t) = (\sigma(u_1(t)), \ldots, \sigma(u_n(t)))$ is the internal state of the output units of the network.
\end{thm}
Backpropagation through automatic differentiation is memory intensive due to the need for values across every point in time to be recalculated at each update of the hidden state. Therefore, gradients are instead computed numerically at the final time step, then, through another ODE solver such as the Adjoint method \cite{Adjoint}, the gradients at all points in time are calculated. In cases such as with Neural ODEs and LNNs \cite{NODES, LNNS}, it is possible to derive closed-form gradient calculations by solving a system of differential equations. This method of backpropagation is less memory intensive and more efficient as the Adjoint Method in Neural ODEs uses \( \mathcal{O}(1) \) memory \cite{NODES}.

\subsection{Liquid Neural Networks}
Liquid neural networks \cite{LNNS} were formulated through a neuron-spiking biological process. This process was integrated into the development of CTRNNs, allowing them to be more dynamic and providing benefits similar to NODEs \cite{NODES}. Formally, the hidden state of these liquid networks is formulated as follows:   
\begin{equation}
\frac{d\boldsymbol{x}(t)}{dt} = -\left[\frac{1}{\tau} + f(\boldsymbol{x}(t), \boldsymbol{I}(t), t, \theta)\right] \boldsymbol{x}(t)+ f(\boldsymbol{x}(t), \boldsymbol{I}(t), t, \theta) A,
\end{equation}
where $A$ is a parameter and $\tau$ denotes a time constant that aids the network in maintaining a numerically stable state.

Despite the formulation's resemblance to NODEs and CTRNNs, the adjoint method is not used for gradient calculation here since the differential equation that governs the hidden state is a set of stiff differential equations leading to an exponential increase of discretization steps when numerically simulating the equation using a Runge-Kutta-based integrator \cite{Press}. Hence, a new type of numerical integrator is necessary, for the adjoint method commonly implements a Runge-Kutta-based Dortmund Prince integrator \cite{NODES,LNNS}.  \cite{LNNS} introduces the Fused solver, a combination of the implicit and explicit Euler Solver. This fused solver discretizes the interval $[0,T]$ as 
\begin{equation}
\mathbf{x}(t + \Delta t) = \frac{\mathbf{x}(t) + \Delta t  f(\mathbf{x}(t), \boldsymbol{I}(t), t, \theta)A}{1 + \Delta t \left(\frac{1}{\tau} + f(\mathbf{x}(t), \boldsymbol{I}(t), t, \theta)\right)}
\end{equation}

Backpropagation through time can be applied as the solver is unrolled throughout time, streamlining the gradient calculation.
These liquid networks offer enhanced expressiveness compared to Neural ODEs and CTRNNs due to the increased complexity of the hidden state. The theorems below demonstrate that the proposed LTC in \cite{LNNS} satisfies a universal approximation property (Theorem 2.4) and that the time constant and the state of the neurons remain within a finite range (Theorem 2.5):
\begin{thm}
Let \( x \in \mathbb{R}^n \), \( S \subset \mathbb{R}^n \), and \( \dot{x} = F(x) \) be an autonomous ODE with \( F : S \to \mathbb{R}^n \) a \( C^1 \)-mapping on \( S \). Let \( D \) denote a compact subset of \( S \) and assume that the simulation of the system is bounded in the interval \( I = [0, T] \). Then, for a positive \( \epsilon \), there exists an LTC network with \( N \) hidden units, \( n \) output units, and an output internal state \( u(t) \), described by Eq. 1, such that for any rollout \( \{ x(t) \mid t \in I \} \) of the system with proper network initialization of initial value \( x(0) \in D \),

\[
\max_{t \in I} |x(t) - u(t)| < \epsilon.
\]
\end{thm}

\begin{thm}
Let \( x_i \) denote the state of neuron \( i \) within an LTC, identified by Eq. 1, and let neuron \( i \) receive \( M \) incoming connections. Then, the hidden state of any neuron \( i \), on a finite interval \( \text{Int} \in [0, T] \), is bounded as follows:
\[
\min(0, A_{i}^{\text{min}}) \leq x_i(t) \leq \max(0, A_{i}^{\text{max}})
\]
\end{thm}
\subsection{Quantum Computing}
  Quantum computing leverages the principles of quantum mechanics in order to perform operations beyond the scope of classical computers. The necessity for quantum computers is more practically illustrated by Moore's Law, which states that the number of transistors in an integrated circuit will double approximately every other year, continuously reducing the size of transistors in circulation. When dealing with objects as small as our current transistors, quantum tunneling phenomena inhibit their functionality. Physical theory in quantum mechanics allows us to model these microscopically small objects and develop more efficient and robust computers. The probabilistic nature of quantum mechanical theory offers additional benefits when applied to building computers that classical mechanics theory does not. At the backbone of all of quantum mechanical theory is the Schrodinger equation \cite{Schrodinger}, which states that for a given quantum mechanical system, its wave equation follows the partial differential equation below:\begin{equation}
    i\hbar\frac{\partial}{\partial t}\Psi (x,t) = \frac{-h}{2m}\nabla^2\Psi(x,t)+V(x,t)\Psi(x,t)
\end{equation}
where 
\begin{itemize}
    \item $i$ is the imaginary constant
    \item $\hbar$ is Planck's constant
    \item $\Psi(x,t)$ denotes the wave function of the system
    \item $\nabla^{2}$ denotes the Laplacian of the system
    \item $V(x,t)$ is the potential energy of the system
\end{itemize}
In most applications of quantum computing, the quantum mechanical system is realized by tracking the evolution of atoms over time. The superposition of these systems is critical for quantum computing as operations can be done to arbitrary states of the system simultaneously without physically applying the transformation to all possible states of the system. The quantum analog of the classical bit is the qubit. Physically, it is realized through a quantum mechanical system in which certain operations may be applied to the systems, \emph{i.e., a rotation along an axis}, without knowing the state of the system. While the theory of quantum computing is based on quantum mechanics, the details generally rely on the analysis of the qubit. \\
 \\While a qubit can be in a superposition of states, upon measurement, it must collapse to a basis state, which we label $|0\rangle$ or $|1\rangle$. It is impossible to calculate deterministically which state a qubit will collapse into. We can, however, determine the probabilities for each possibility through the wave function. Quantum machine learning aims to alter these probabilities by altering the state of the function. Through a series of operations in which the parameters are controlled, it is possible to create a state where a qubit will necessarily collapse into a desired state upon measurement.
\\Mathematically, a qubit is an element of a Hilbert space. In order to build a composite system of qubits, it is necessary to first derive the notion for a single qubit system. Let $\mathcal{H}$ denote a Hilbert Space with basis vectors $$|0\rangle  = \begin{bmatrix}
    1\\
    0
    
\end{bmatrix} \: and \;\; |1\rangle = \begin{bmatrix}
    0\\
    1
\end{bmatrix}$$
\\\ An element $x\in\mathcal{H}$ is a vector with coordinates$$x = \begin{bmatrix}
    x_1\\
    x_2
\end{bmatrix}$$

\begin{figure}
    \centering
    \includegraphics[scale = 0.15]{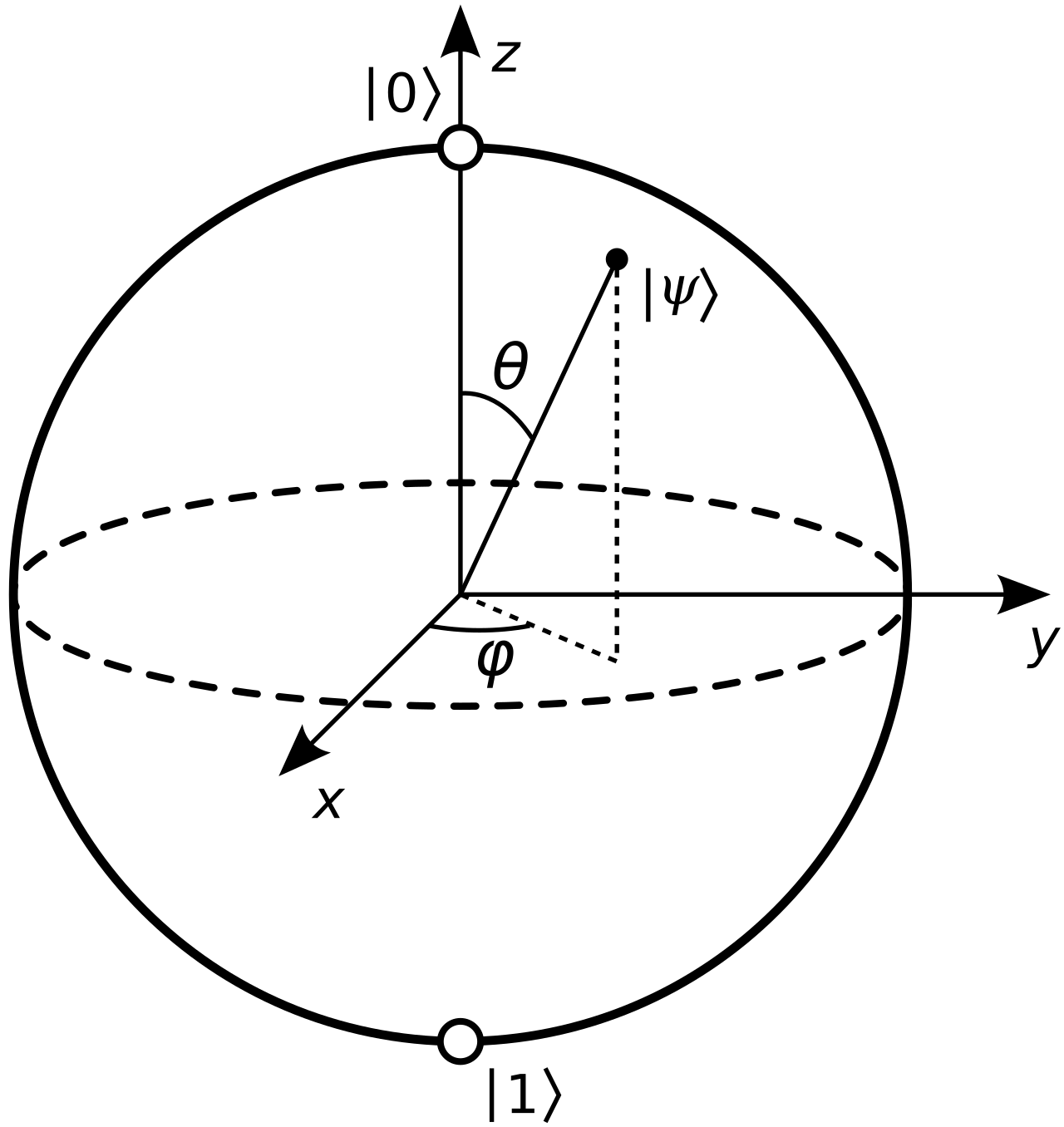}
    \caption{Bloch Sphere}
    \label{fig:enter-label}
\end{figure}

where $x_{1}, x_{2} \in \mathbb{C}$, such that $|x_{1}|^{2}+|x_{2}|^{2} = 1$. These coordinates represent the probability of the qubit collapsing into each respective state upon measurement. A composite system of qubits is formed through the tensor product we define as

\[
|\psi \phi\rangle = |\psi\rangle \otimes |\phi\rangle
\]

where

\[
|\psi\rangle = \begin{bmatrix}
    \psi_1 \\
    \psi_2
\end{bmatrix}, \quad
|\phi\rangle = \begin{bmatrix}
    \phi_1 \\
    \phi_2
\end{bmatrix}
\]

Then, the tensor product is given by

\[
|\psi \phi\rangle = \begin{bmatrix}
    \psi_1 \\
    \psi_2
\end{bmatrix} \otimes \begin{bmatrix}
    \phi_1 \\
    \phi_2
\end{bmatrix} = \begin{bmatrix}
    \psi_1 \phi_1 \\
    \psi_1 \phi_2 \\
    \psi_2 \phi_1 \\
    \psi_2 \phi_2
\end{bmatrix}
\]

Additionally, we define the outer product as 

\[
|\psi\rangle \langle \psi| = \begin{bmatrix}
    \psi_1 \\
    \psi_2
\end{bmatrix}
\begin{bmatrix}
    \psi_1^* & \psi_2^*
\end{bmatrix}
= \begin{bmatrix}
    \psi_1 \psi_1^* & \psi_1 \psi_2^* \\
    \psi_2 \psi_1^* & \psi_2 \psi_2^*
\end{bmatrix}
\]

Finally, we define the inner product as 

\[
\langle \psi | \psi \rangle = \psi_1^* \psi_1 + \psi_2^* \psi_2
\]
As qubits are formulated mathematically through a vector with coordinates, we are able to leverage linear algebraic operations including rotation operators and linear operators. Examples of such operators are produced below.
\begin{figure}[h!]
    \centering
    \begin{tabular}{c@{\hskip 1.5cm}c@{\hskip 1.5cm}c@{\hskip 1.5cm}c@{\hskip 1.5cm}c}
        % Hadamard Gate
        \begin{tikzpicture}
            \node at (0, 0) {$H = \frac{1}{\sqrt{2}}\begin{pmatrix}
                1 & 1 \\
                1 & -1
            \end{pmatrix}$};
        \end{tikzpicture} &
        % Pauli X Gate
        \begin{tikzpicture}
            \node at (0, 0) {$X = \begin{pmatrix}
                0 & 1 \\
                1 & 0
            \end{pmatrix}$};
        \end{tikzpicture} &
        % Pauli Y Gate
        \begin{tikzpicture}
            \node at (0, 0) {$Y = \begin{pmatrix}
                0 & -i \\
                i & 0
            \end{pmatrix}$};
        \end{tikzpicture} &
        % Pauli Z Gate
        \begin{tikzpicture}
            \node at (0, 0) {$Z = \begin{pmatrix}
                1 & 0 \\
                0 & -1
            \end{pmatrix}$};
        \end{tikzpicture} &
        % CNOT Gate
        \begin{tikzpicture}
            \node at (0, 0) {$\text{CNOT} = \begin{pmatrix}
                1 & 0 & 0 & 0 \\
                0 & 1 & 0 & 0 \\
                0 & 0 & 0 & 1 \\
                0 & 0 & 1 & 0
            \end{pmatrix}$};
        \end{tikzpicture} \\
        \text{Hadamard Gate (H)} & \text{Pauli-X Gate (X)} & \text{Pauli-Y Gate (Y)} & \text{Pauli-Z Gate (Z)} & \text{CNOT Gate}
    \end{tabular}
\end{figure}

\begin{figure}[H]
    \centering
    \includegraphics[scale = 1.15]{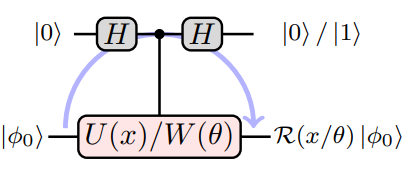}
    \caption{\label{fig:Quantum Residual Block}The Quantum Residual Block}
\end{figure}

\newpage
\section{Theoretical Framework}

\cite{QRESNET} proposed a new framework for a quantum residual block in which they define a quantum residual operator applied to the entangled state $\ket{\varphi_{0}}\otimes |0\rangle$. We utilize this approach to derive a quantum liquid network.  A traditional residual block encoded into a quantum neural network is of the form:
\begin{equation}
\ket{\psi_{t+1}} = \ket{\psi_{t}} + \mathcal{U}\ket{\psi_{t}}.
\end{equation}
This approach does not retain the benefits of quantum computing. Instead, the authors of \cite{QRESNET} present a quantum residual block that leverages the probabilistic nature of quantum computing to further enhance the capability of a residual block. Let $ \mathcal{H}_{1}= \mathbb{C}^{2^n}$ and $ \mathcal{H}_{0}= \text{span}\left\{ \ket{0} \right\}$ denote Hilbert spaces. Consider an initial quantum state $\ket{\varphi_{0}}\in \mathcal{H}_{1}$ and an auxiliary qubit $\ket{0}\in\mathcal{H}_{0}$ as illustrated in Figure \ref{fig:Quantum Residual Block}. The implementation of the quantum residual operator is realized in the subspace of the ancilla qubit. Once the residual operator is applied to the initial state, one can then obtain the results within the subspace $V = \left\{ |x\rangle\langle x| \mid |x\rangle \in \mathcal{H}_{0} \right\}$
 \cite{QRESNET}.

Inspired by \cite{QRESNET}, we develop an LQNet and a CTRQNet. First, we apply either a Hadamard gate to $\ket{0}$ or a corresponding transformation to the state $\ket{\varphi_{0}}$ depending on whether we are in the encoding stage or parameterized stage. We then apply a CNOT gate to the qubits using $\ket{0}$ as the control qubit and $\ket{\varphi_{0}}$ as the target qubit. From there, we apply another Hadamard to the quantum state whose output is measured in the auxiliary qubit subspace $\mathcal{H}_{0}$. The result of this operator is then added to the quantum state lying in the subspace of the first qubit. A final operation is performed on the state lying in $\mathcal{H}_{0}$ to realize the residual connection. Rewriting in a more succinct notation, we propose a Quantum Liquid Neural Network as follows :

Consider the entangled state $\ket{\varphi_{0}}\otimes \ket{0}$. Then 
the output of the residual connection  in the encoding stage(Eq. 3.1) can be rewritten in the following form:
\begin{equation}
F(\ket{\psi_{t}}) \triangleq \text{tr}_{\mathcal{H}_{0}}\braket{\psi_t|F^{\dagger}(\ket{\varphi_{0}}\otimes \ket{0})F|\psi_t},
\end{equation}
where $F:\mathcal{H}_{1}\rightarrow V$ denotes the action of the quantum residual block acting on classical data $\ket{x}\in \mathcal{H}_{0}$ mapping to the subspace $ V= \left\{ |x\rangle\langle x| \mid \ket{x} \in \mathcal{H}_{0} \right\}$. We can think of the mapping $F$ as a sequence or composition of unitary operators 
\begin{equation}
 F \triangleq (H \otimes U)\circ (\text{CNOT})\circ (H \otimes I),
\end{equation}
where for the initial state $\ket{\varphi_{0}}$, $U \triangleq \ket{\varphi_{0}}\bra{\varphi_{0}}$, $\text{CNOT}$ denotes the controlled-not gate, $H$ denotes the Hadamard gate, and $I$ denotes the identity matrix. Note that $F$ is unitary by construction. Similarly, we define $\tilde{F}:\mathcal{H}_{1}\rightarrow V$ by 
\begin{equation}
\tilde{F} \triangleq (H \otimes W(\theta)) \circ (\text{CNOT}) \circ (H \otimes I),
\end{equation}
where, as in \cite{QRESNET}, we replace the encoding operator with the parameterized gate $W(\theta)$ while the remaining gates remain unchanged. Let 
\begin{equation}
\tilde{F}(\ket{\psi_{t}},\theta) = \text{tr}_{\mathcal{H}_{0}}\braket{\psi_{t}|\tilde{F}^{\dagger}(\ket{\varphi_{0}}\otimes \ket{0})\tilde{F}|\psi_{t}},
\end{equation}
where $\ket{\psi_{t}}$ denotes the evolution of initial quantum state $|\varphi_{0}\rangle$  on layer $t$.
%then we have that 
%\begin{equation}
%\mathcal{Q}(\ket{\varphi_{0}},\theta) = f\braket{\psi_{t}} + F\braket{\psi_{t},\theta}
%\end{equation}
${\tilde{F}(\ket{\psi_{t}},\theta)}$ is equivalent to $W(\theta)\ket{\varphi_{0}}$, in $W(\theta)\ket{\varphi_{0}} \triangleq \mathcal{W}(\theta)$. With our definition of the residual block $\tilde{F}(\psi_{t},\theta)$, we have that in the limit, a model equation for quantum NODEs as follows
\begin{equation}
\frac{d{\varphi}(t)}{dt} = {\tilde{F}(\ket{\psi_{t}},\theta)},
\end{equation} 
where $\varphi(t)$ denotes a continuous function in time, $t$. The solution of Eq.(3.6), when evaluated at $t=T$, forms the output of the quantum NODE block. Substitution of this into the equation for liquid networks, we have that
\begin{equation}
\frac{d{\varphi}(t)}{dt} = -\bigg[\tau^{-1}+{\tilde{F}(\ket{\psi_{t}},\theta)}\bigg]\varphi(t)+{\tilde{F}(\ket{\psi_{t}},\theta)}
\end{equation}

These new equations form our newly created model, hereafter referred to as LQNets. Utilizing the Fused Solver developed by (Hasani et al., 2020) \cite{LNNS}, we simulate the differential equation to obtain 
\begin{equation}
    {\varphi}(t+\Delta t) = \frac{{\varphi}(t)+\Delta t {\tilde{F}(\ket{\psi_{t}},\theta)}}{1+\Delta t \big(\tau^{-1}+{\tilde{F}(\ket{\psi_{t}},\theta)}\big)}
\end{equation} 

Following a similar argument, we develop a quantum time-continuous recurrent neural networks (CTRQNet). Recall that a CTRNN is defined by the equation 
\[\frac{d\textbf{x}(t)}{dt}= -\frac{1}{\tau}\textbf{x}(t)+W\sigma(\textbf{x}(t)),\]
in which $\sigma:\mathbb{R}\rightarrow (0,1)\in C^{1}$ is a sigmoid function in accordance with Theorems 2.2 and 2.3, and $W$ denotes a weight matrix. With Eq. (3.6) defined, we can amend the standard CTRNN equation to read 
\begin{equation}
    \frac{d{\varphi}(t)}{dt}= -\tau^{-1}\varphi(t)+{\tilde{F}(\ket{\psi_{t}},\theta)}, 
\end{equation}
where the quantity ${\tilde{F}(\ket{\psi_{t}},\theta)}$ replaces $W(\sigma(\textbf{x}(t)))$ and is equivalent to the internal state of the output units of the network (cf. Theorem 2.3). This new equation (Eq. 3.9) forms the proposed model for CTRQNets.
\section{Results}
In this section we implement the 2 newly created models, the LQNet and CTRQNet. In 4.1 we discuss the methodology used throughout the experiments, in 4.2 we discuss the results of our experiments on each of the datasets listed.
\subsection{Methodology}
We perform tests of our networks using the following benchmark datasets: 
\begin{itemize}
    \item MNIST Binary Classification
    \item FMNIST Binary Classification
    \item Wisconsin Breast Cancer Dataset
    \item CIFAR 10 Binary Classification
    \item CIFAR 10 Binary Classification with Resnet-151 as downsampling layers
\end{itemize}
Throughout all experiments, we use a 3-qubit quantum circuit to simulate the LQNet and CTRQNet and four discretization steps with a step length of 0.1. Due to the computational complexity of pseudo-quantum computing and high training times, multi-class classification is not feasible within our time constraints. However, a functional quantum computer can make this possible. We implement these algorithms in Qiskit and Pytorch.
\subsection{Baseline Testing}
\subsubsection{MNIST}
The MNIST dataset consists of grayscale images of handwritten digits from 0 to 9. Our models use 0s and 1s as binary classification. The training set consists of 150 images, the validation set 50 images, and the testing set 2115 images. We present our loss curves below.

In Figure 4, both the LQNet and CTRQNet are able to converge to and maintain performance around an optimal state. Both models' losses approach 0, and the lost curves exhibit no volatility. While the QNN also converges to an optimal state, its minimum loss is roughly 0.3, while the LQNet and CTRQNet achieve a loss of near 0.001. Additionally, the LQNet and CTRQNet reached optimal states in 10 steps, whereas  QNNs were not able to reach an optimal state within 50 steps.

In Figure 5, the LQNet and CTRQNet achieve maximal performance in terms of accuracy on the validation set after 10 steps. Both models can maintain optimal performance consistently after first achieving it, as was observed in the loss curves for both models. The QNN takes longer than these models to achieve an optimal state, requiring around 20 steps, but still maintains the consistency seen in the other two networks.
\begin{figure}[H]
    \centering
    \subfigure[LQNet Loss Curve]{%
        \includegraphics[width=0.40\textwidth]{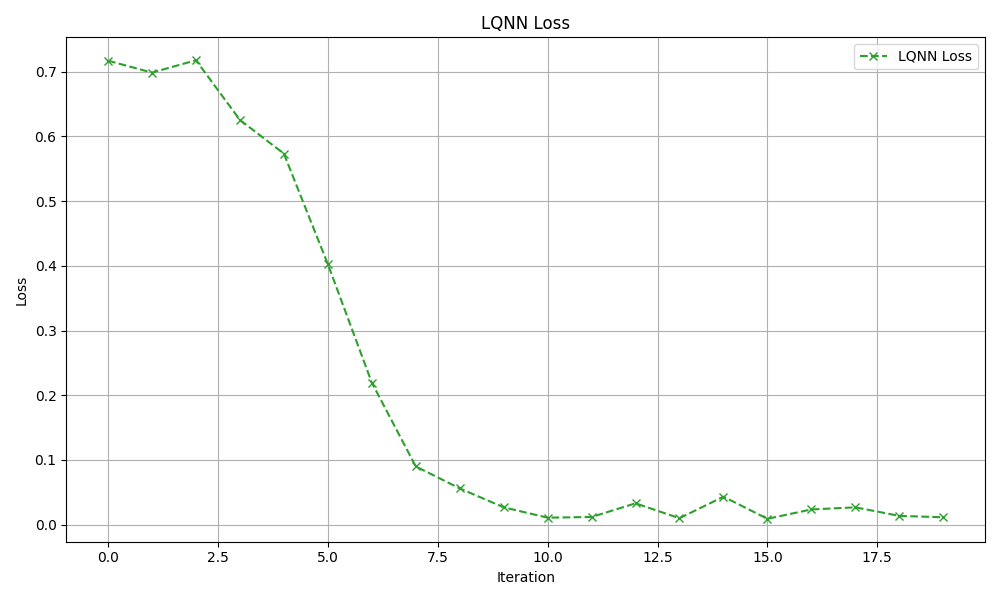}
        \label{fig:LQNetLossMNIST}
    }
    \subfigure[CTRQNet Loss Curve]{%
        \includegraphics[width=0.40\textwidth]{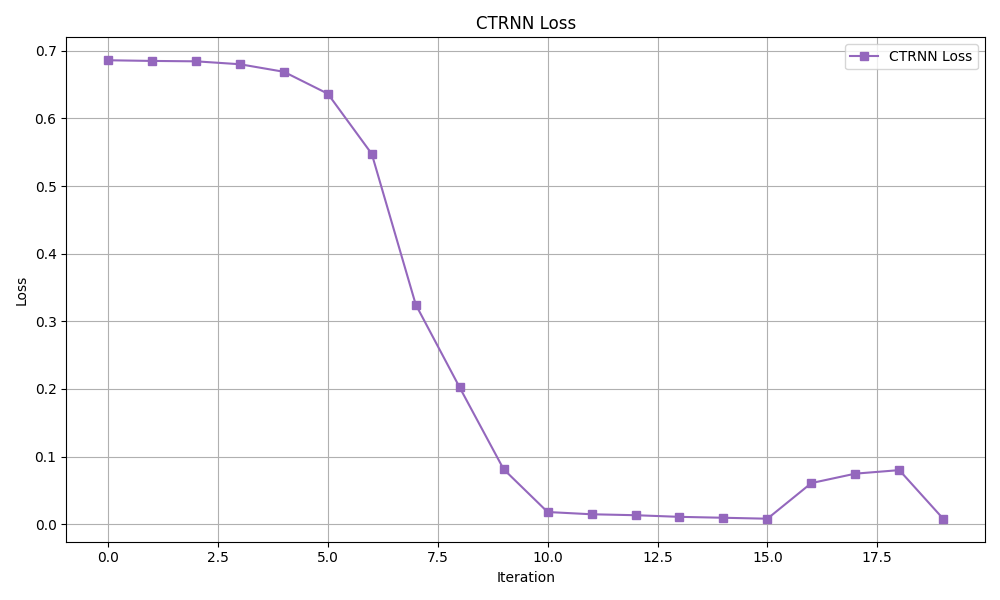}
        \label{fig:CTRQNetLossMNIST}
    } \\
    \subfigure[QNN Loss Curve]{%
        \includegraphics[width=0.40\textwidth]{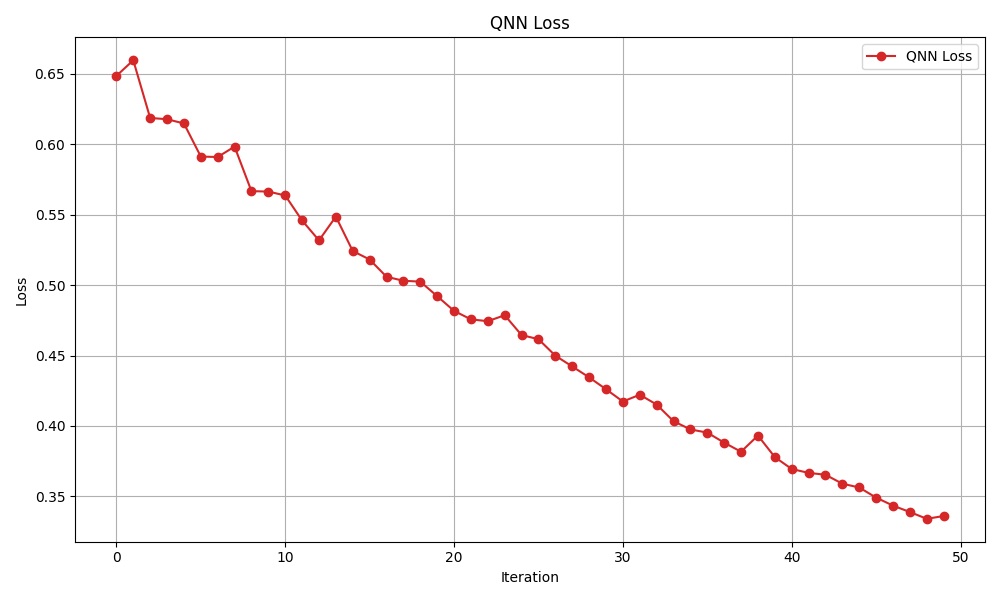}
        \label{fig:QNetLossMNIST}
    }
    \subfigure[MNIST Loss Comparison]{%
        \includegraphics[width=0.40\textwidth]{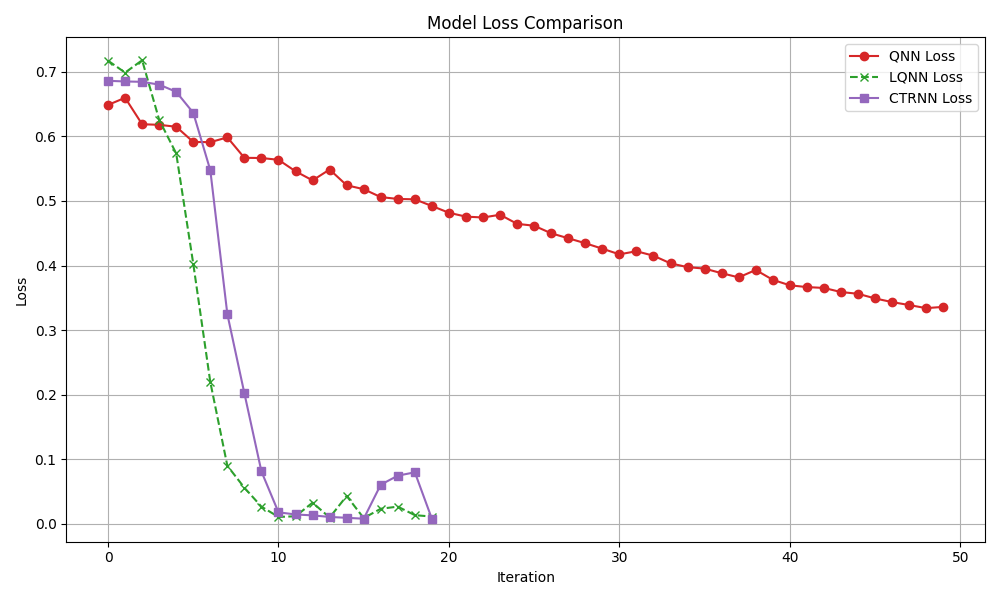}
        \label{fig:LossComparisonMNIST}
    }
    \caption{Loss Curves and Comparisons for Quantum Networks on MNIST}
\end{figure}

\begin{figure}[H]
    \centering
    \subfigure[LQNet Accuracy Curve]{%
        \includegraphics[width=0.40\textwidth]{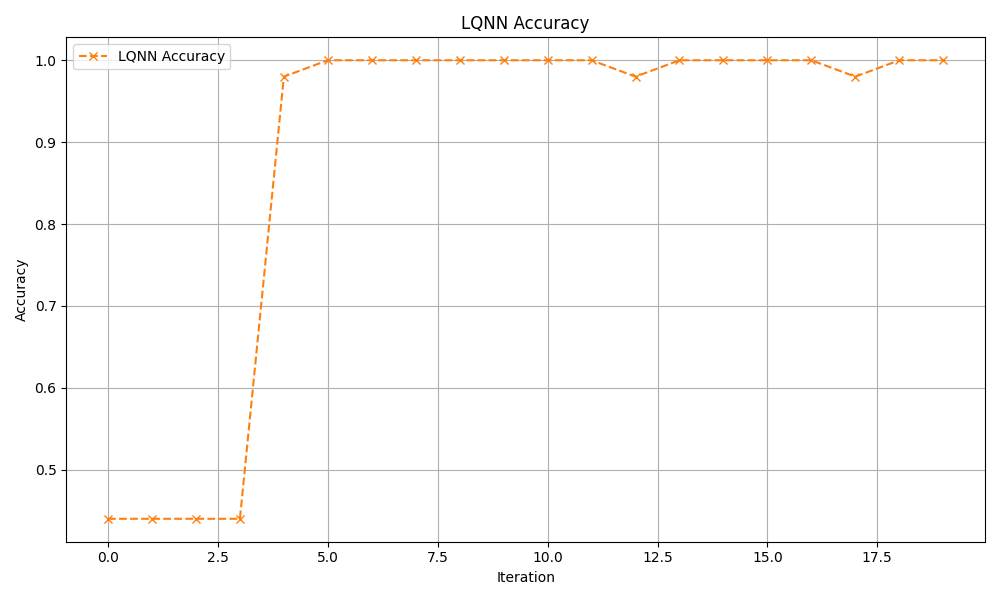}
        \label{fig:LQNetAccuracyMNIST}
    }
    \subfigure[CTRQNet Accuracy Curve]{%
        \includegraphics[width=0.40\textwidth]{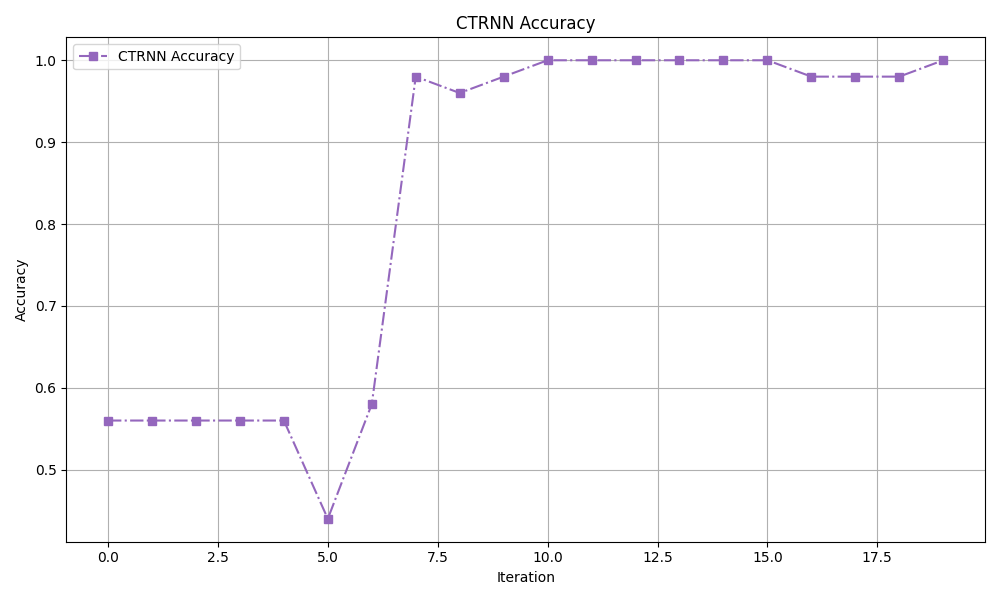}
        \label{fig:CTRQNetAccuracyMNIST}
    }\\
    \subfigure[QNN Accuracy Curve]{%
        \includegraphics[width=0.40\textwidth]{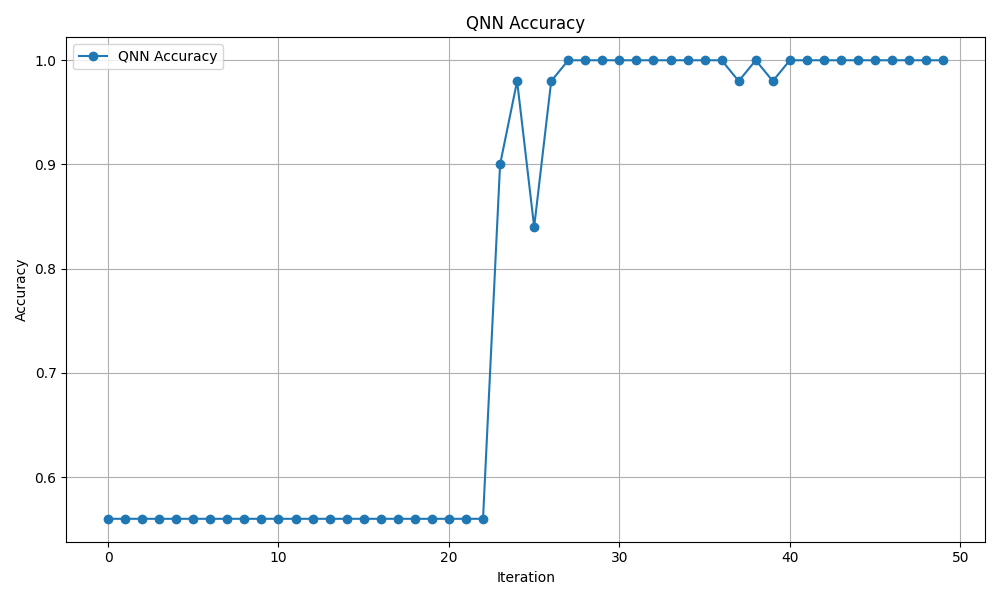}
        \label{fig:QNetAccuracyMNIST}
    }
    \subfigure[MNIST Accuracy Comparison]{%
        \includegraphics[width=0.40\textwidth]{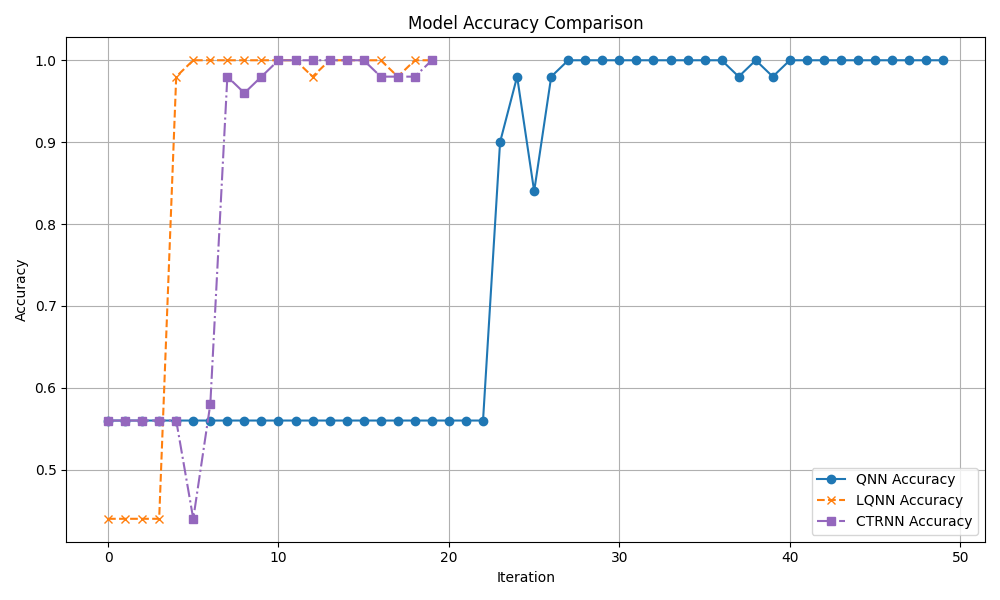}
        \label{fig:AccuracyComparisonMNIST}
    }
    \caption{Accuracy Curves and Comparisons for Quantum Networks on MNIST}
\end{figure}

\begin{table}[H]
    \centering
    \resizebox{\textwidth}{!}{
        \begin{tabular}{|l|c|c|c|c|c|c|}
            \hline
            \textbf{Model} & \textbf{Accuracy (\%)} & \textbf{F1 Score} & \textbf{Precision Class 1 (\%)} & \textbf{Precision Class 0 (\%)} & \textbf{Recall Class 1 (\%)} & \textbf{Recall Class 0 (\%)} \\
            \hline
            CTRQNet & 99.81 & 99.81 & 100 & 100 & 100 & 100 \\
            QLNet   & 99.67 & 99.67 & 99  & 100 & 100 & 99  \\
            QNN    & 99.53 & 99.53 & 99  & 100 & 100 & 99  \\
            \hline
        \end{tabular}
    }
    \caption{Performance Metrics of Different Models}
    \label{tab:performance_metrics_mnist}
\end{table}
According to Table 1, all models reached at least 99\% accuracy. However, on more challenging tasks such as CIFAR 10, the QNN is unable to learn due to its rigid structure, whereas the LQNet and CTRQNet are able to converge rapidly due to their continuous hidden states.

\subsubsection{FMNIST}
We analyze our models on the Fashion MNIST dataset through binary classification, using 150 training images, 50 validation images, and 2000 testing images. Every 15 steps in the optimization process, we report the model's performance on the validation dataset, as shown below.

\begin{figure}[H]
    \centering
    \subfigure[LQNet Loss Curve]{%
        \includegraphics[width=0.45\textwidth]{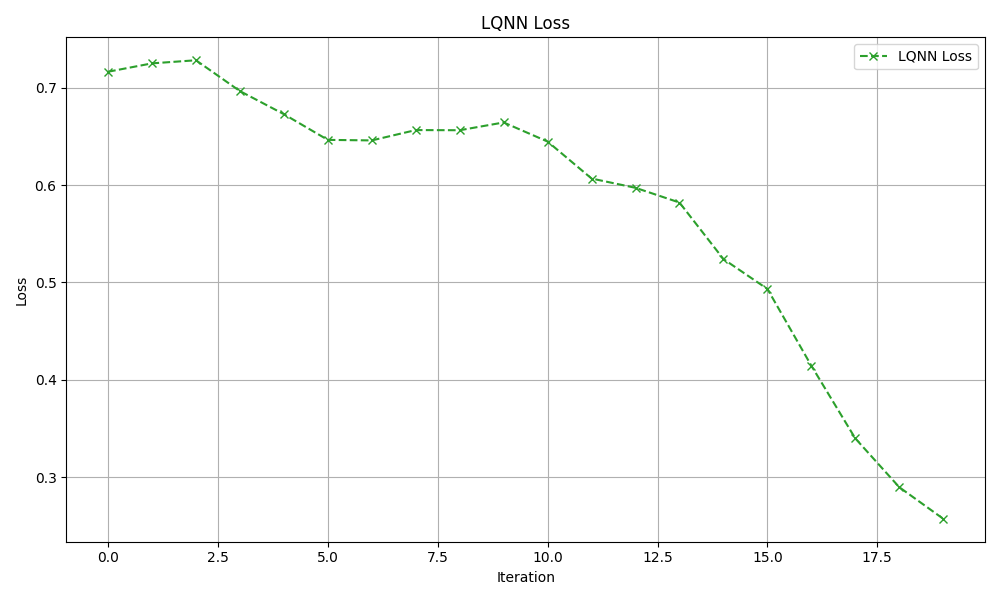}
        \label{fig:LQNetLossFMNIST}
    }
    \subfigure[CTRQNet Loss Curve]{%
        \includegraphics[width=0.45\textwidth]{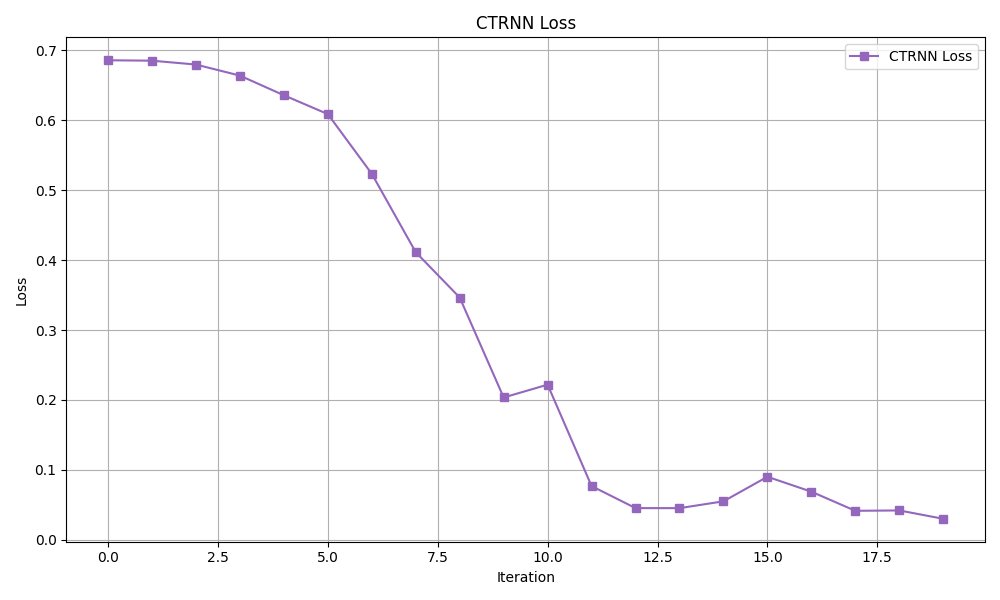}
        \label{fig:CTRQNetLossFMNIST}
    }\\
    \subfigure[QNN Loss Curve]{%
        \includegraphics[width=0.45\textwidth]{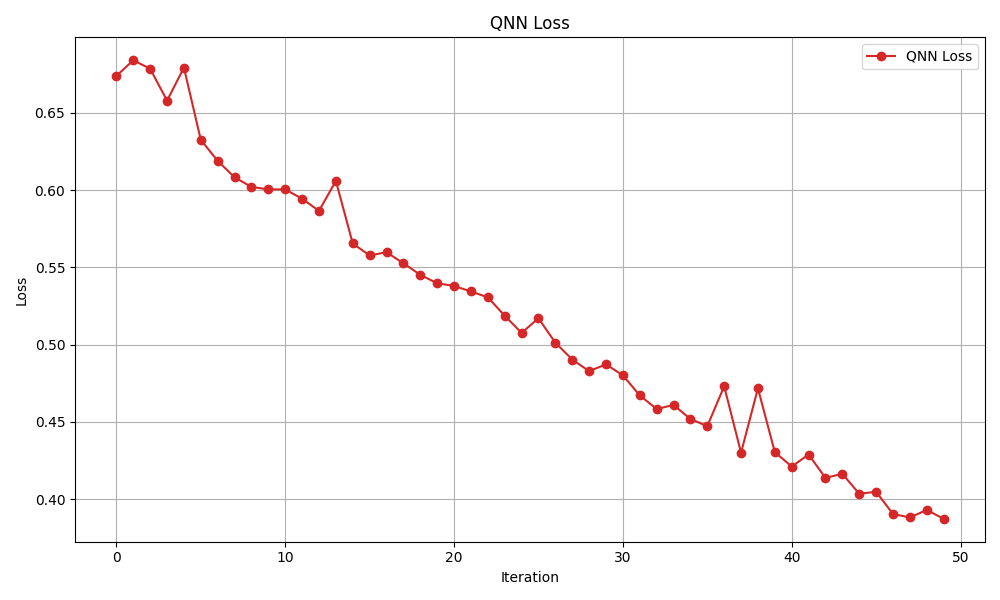}
        \label{fig:QNetLossFMNIST}
    }
    \subfigure[FMNIST Loss Comparison]{%
        \includegraphics[width=0.45\textwidth]{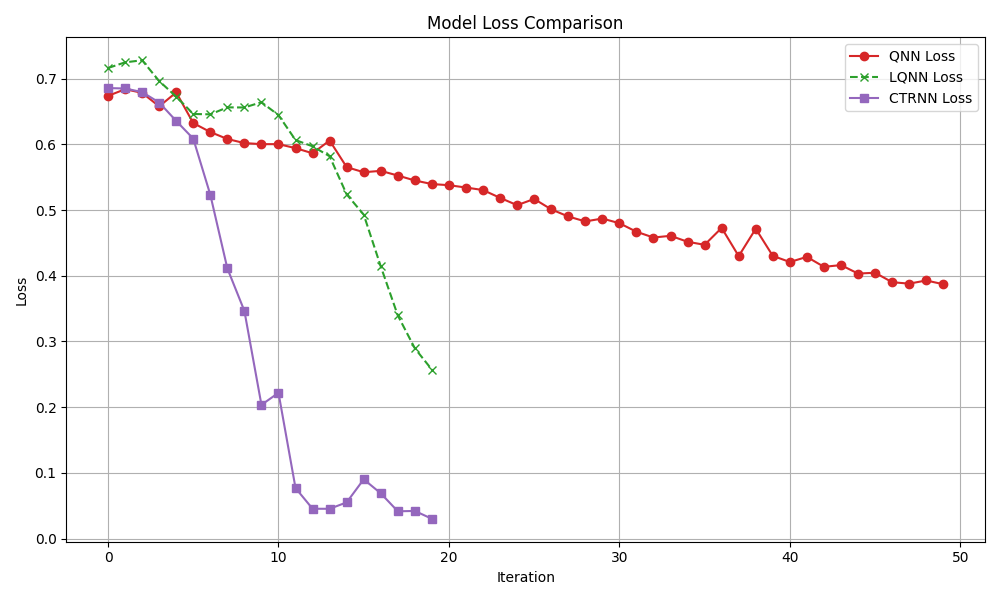}
        \label{fig:LossComparisonFMNIST}
    }
    \caption{Loss Curves and Comparisons for Quantum Networks on Fashion MNIST}
\end{figure}
In Figure 6, the CTRQNet converges to an optimal state within 195 steps, making it the fastest of the three models to reach an optimal state. 
%It has an extremely steep curve, showing that the model is dynamic and can rapidly learn patterns in the data. 
On the other hand, the LQNet reaches an optimal state after 300 steps; it may benefit from more training time as it has not fully converged after 300 steps. The QNN converges to an optimal state in a linear fashion and converges much slower than the LQNet and CTRQNet models.
%Due to the linearity of the curve, we can conclude that the QNN does not enjoy the dynamic learning system that CTRQNetS and LQNetS possess.

\begin{figure}[H]
    \centering
    \subfigure[LQNet Accuracy Curve]{%
        \includegraphics[width=0.45\textwidth]{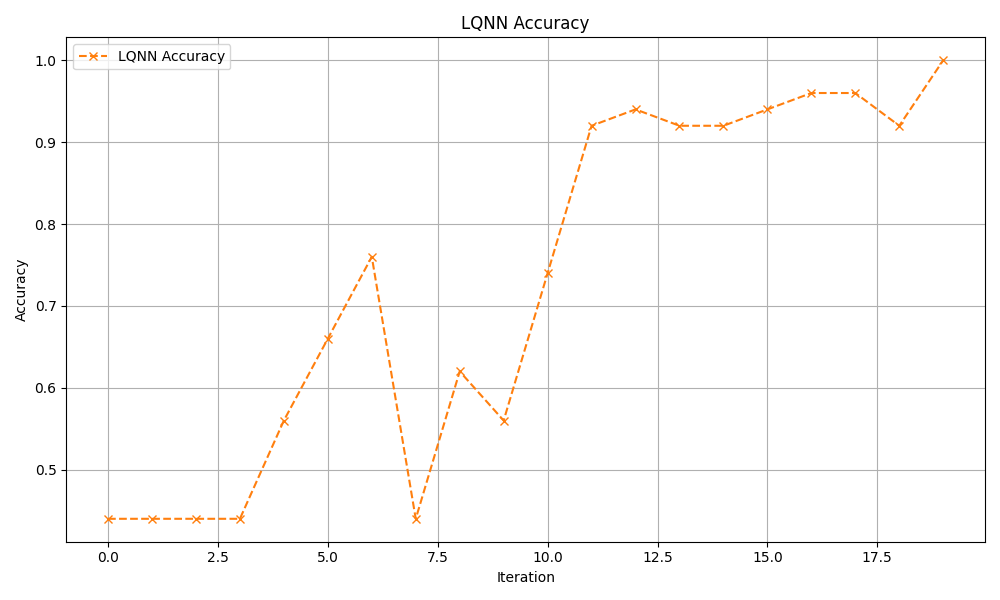}
        \label{fig:LQNetAccuracyFMNIST}
    }
    \subfigure[CTRQNet Accuracy Curve]{%
        \includegraphics[width=0.45\textwidth]{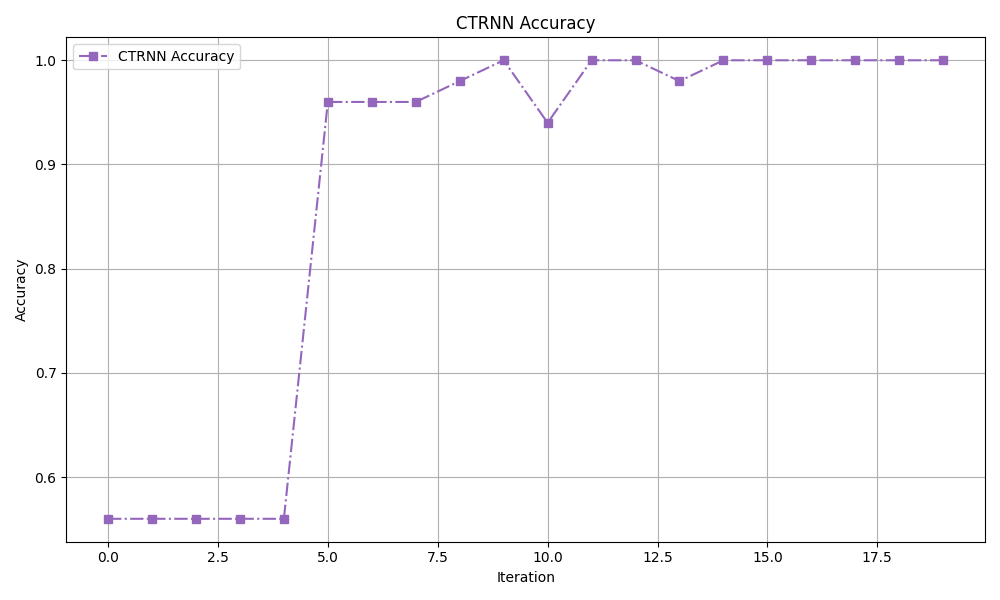}
        \label{fig:CTRQNetAccuracyFMNIST}
    }\\
    \subfigure[QNN Accuracy Curve]{%
        \includegraphics[width=0.45\textwidth]{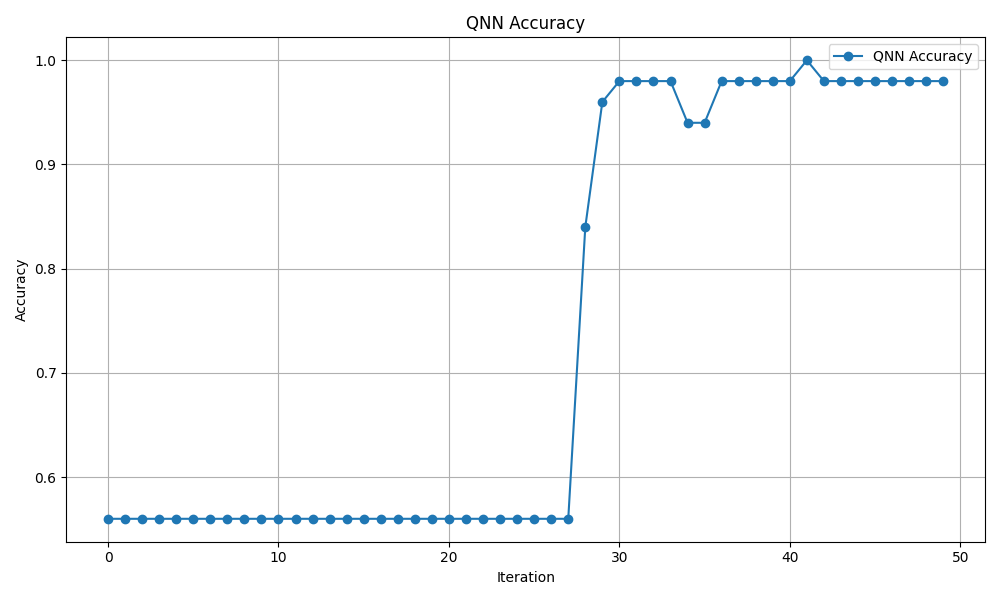}
        \label{fig:QNetAccuracyFMNIST}
    }
    \subfigure[FMNIST Accuracy Comparison]{%
        \includegraphics[width=0.45\textwidth]{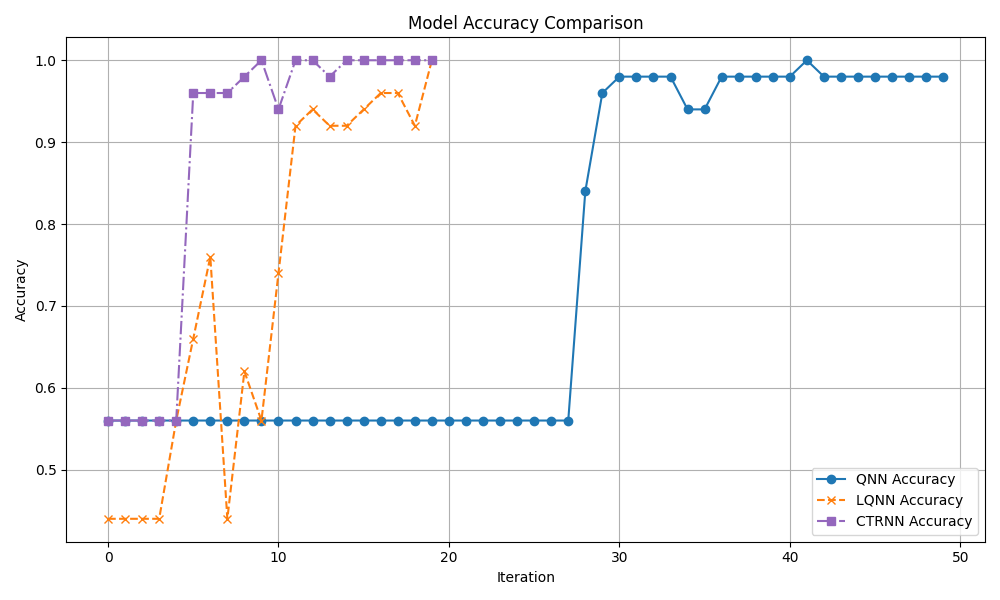}
        \label{fig:AccuracyComparisonFMNIST}
    }
    \caption{Accuracy Curves and Comparisons for Quantum Networks on FMNIST}
\end{figure}
In Figure 7, the accuracy of the CTRQNet shows the most rapid growth of all models. It achieves an optimal state after 125 steps and maintains performance within a reasonable range afterward. We discover through experiments that the CTRQNet is less prone to volatility in its learning process than the LQNet. We also notice that the LQNet displays an increasing accuracy trend but experiences occasional dips throughout the training process, which decrease in size. 
%We attribute this to the dynamic structure of liquid networks, enabling the LQNet to learn much quicker but leading to overfitting when using the classical state of hyperparameters.
%This trend also persists in CTRQNetS, so a more intensive investigation is performed throughout the rest of these baseline tests to analyze this pattern. 
Finally, the QNN takes nearly 405 steps to reach an optimal state and also maintains performance within a reasonable range. Finally, we conclude that the QNN is not dynamic and may not be able to learn as efficiently as its structure is very rigid. Our proposed models, the LQNet and CTRQNet, allow a less rigid structure enabling them to learn more efficiently than QNNs.

\begin{table}[H]
    \centering
    \resizebox{\textwidth}{!}{
        \begin{tabular}{|l|c|c|c|c|c|c|}
            \hline
            \textbf{Model} & \textbf{Accuracy (\%)} & \textbf{F1 Score} & \textbf{Precision Class 1 (\%)} & \textbf{Precision Class 0 (\%)} & \textbf{Recall Class 1 (\%)} & \textbf{Recall Class 0 (\%)} \\
            \hline
            CTRQNet & 99.81 & 99.81 & 100 & 100 & 100 & 100 \\
            QLNet   & 98.40 & 98.40 & 98  & 98  & 98  & 98  \\
            QNN    & 98.15 & 98.15 & 98  & 98  & 98  & 98  \\
            \hline
        \end{tabular}
    }
    \caption{Performance Metrics of Different Models on FMNIST}
    \label{tab:performance_metrics_FMNIST}
\end{table}
According to Table 2, all three models achieved high accuracy, each reaching at least 98\%, but the CTRQNet outperformed the LQNet and QNN models by a slight margin.

%Not only are CTRQNets and LQNets extremely dynamic, they can learn from small datasets and adjust their structure throughout the training process. QNNs do not have these properties as they lack a time-continuous structure. 
We note that the CTRQNet and LQNet converge in 200-250 steps, while QNNs require 405–420 steps. Both models generalize to the entire dataset effectively, given that a training set of 150 images was sufficient to achieve optimal performance on a testing set of 2000 images. While the LQNet converges rapidly, it displays decreasing but significant volatility over the training process. We attribute this to the overfitting problem that arises from using the classical set of hyperparameters on these quantum models. The CTRQNet is observed to be most effective due to its accuracy and speed of convergence, with the LQNet performing in a similar manner.
\subsubsection{Wisconsin Breast Cancer}
The Wisconsin Breast Cancer dataset consists of 405 training samples, 50 validation samples, and 114 testing samples.
\begin{figure}[H]
    \centering
    \subfigure[LQNet Loss Curve]{%
        \includegraphics[width=0.45\textwidth]{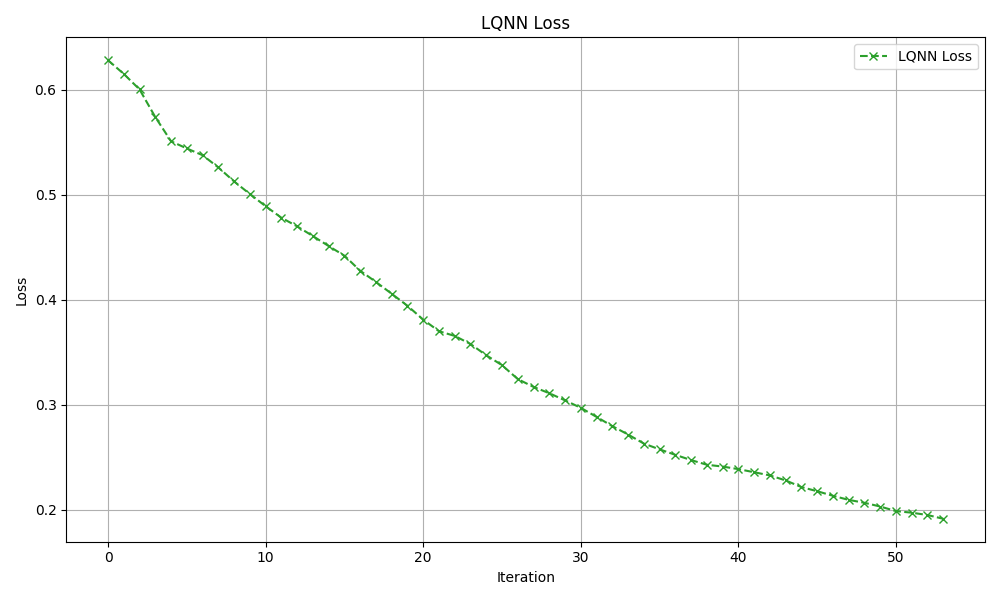}
        \label{fig:LQNetLossWisconsin}
    }
    \subfigure[CTRQNet Loss Curve]{%
        \includegraphics[width=0.45\textwidth]{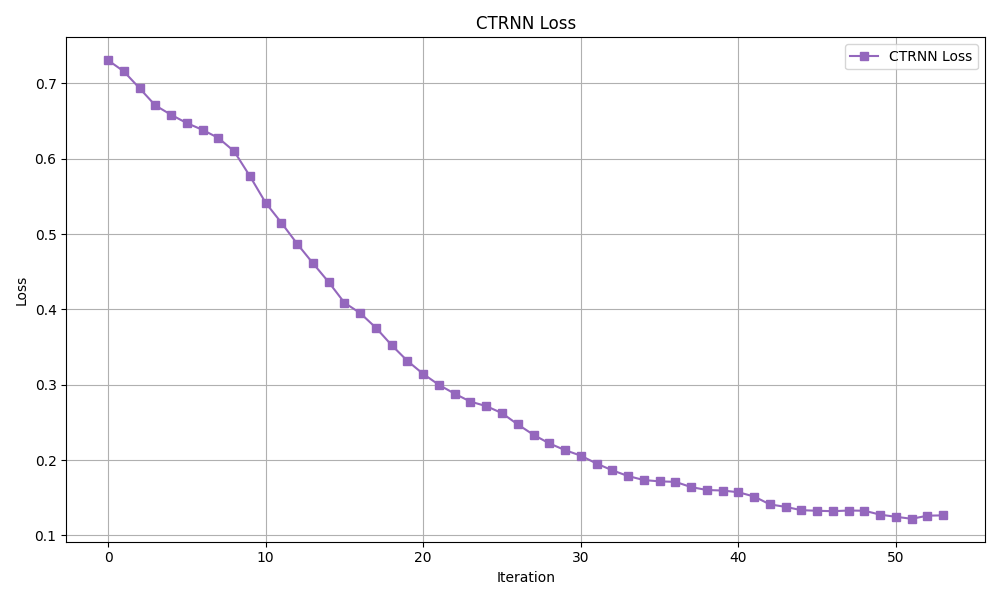}
        \label{fig:CTRQNetLossWisconsin}
    }\\
    \subfigure[QNN Loss Curve]{%
        \includegraphics[width=0.45\textwidth]{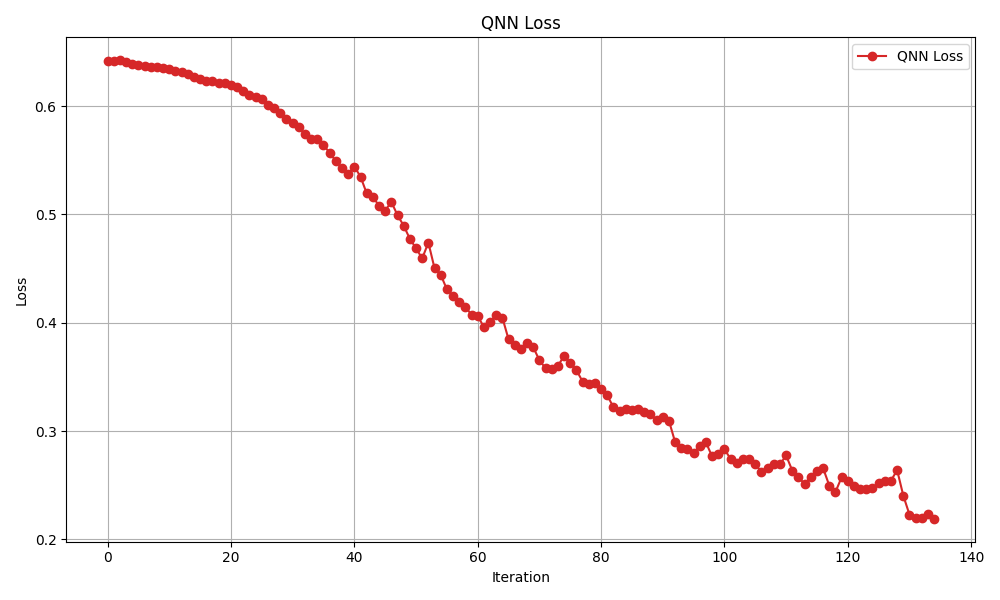}
        \label{fig:QNetLossWisconsin}
    }
    \subfigure[Wisconsin Breast Cancer Loss Comparison]{%
        \includegraphics[width=0.45\textwidth]{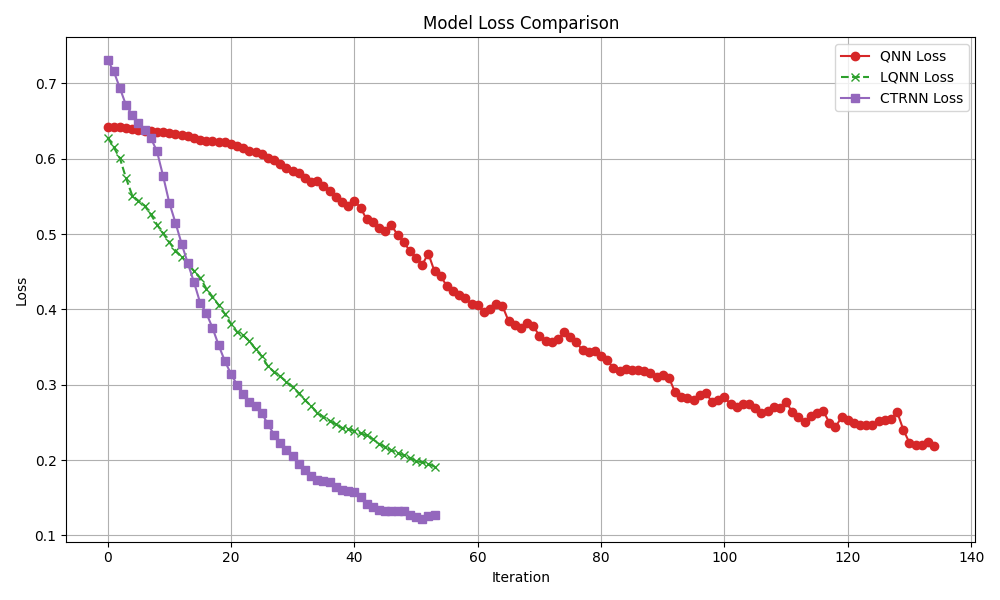}
        \label{fig:LossComparisonWisconsin}
    }
    \caption{Loss Curves and Comparisons for Quantum Networks on Wisconsin Breast Cancer Dataset}
\end{figure}
In Figure 8, all three models show a decreasing loss curve. However, the loss curves of the LQNet and CTRQNet indicate that they converge to an optimal solution much more rapidly than the QNN. The QNN reaches a minimum loss of 0.2, whereas the LQNet and CTRQNet models achieve a loss of 0.03. The QNN also had more training steps than the other two models, having 140 steps compared to the LQNet and CTRQNet 55 steps, respectively.

\begin{figure}[H]
    \centering
    \subfigure[LQNet Accuracy Curve]{%
        \includegraphics[width=0.45\textwidth]{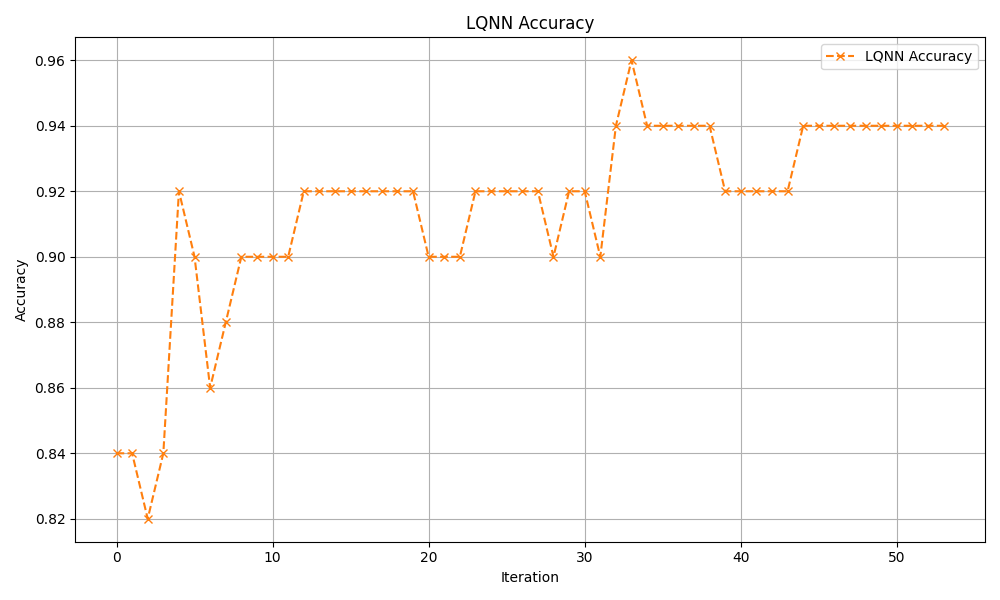}
        \label{fig:LQNetAccuracyWisconsin}
    }
    \subfigure[CTRQNet Accuracy Curve]{%
        \includegraphics[width=0.45\textwidth]{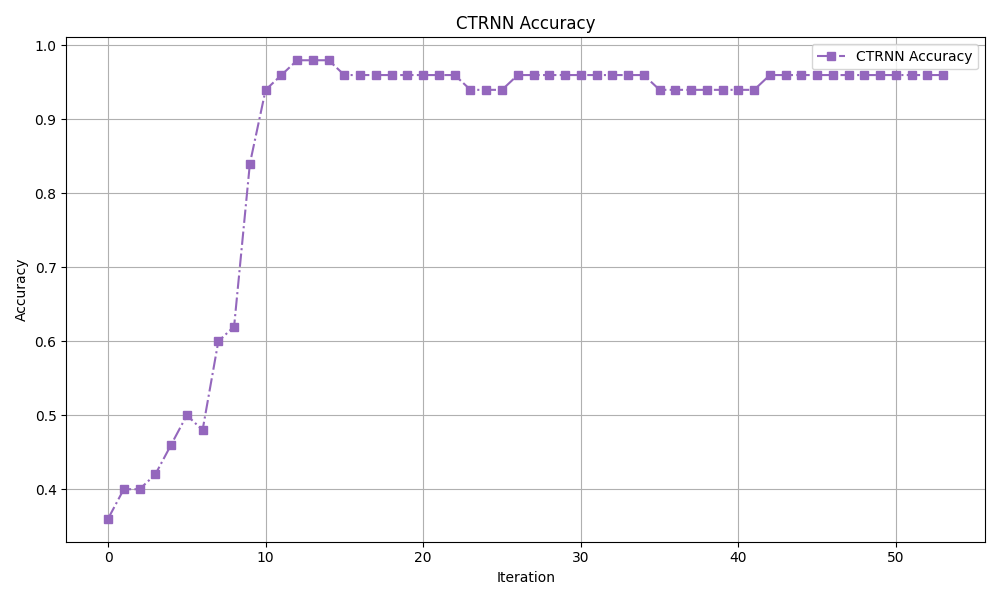}
        \label{fig:CTRQNetAccuracyWisconsin}
    }\\
    \subfigure[QNN Accuracy Curve]{%
        \includegraphics[width=0.45\textwidth]{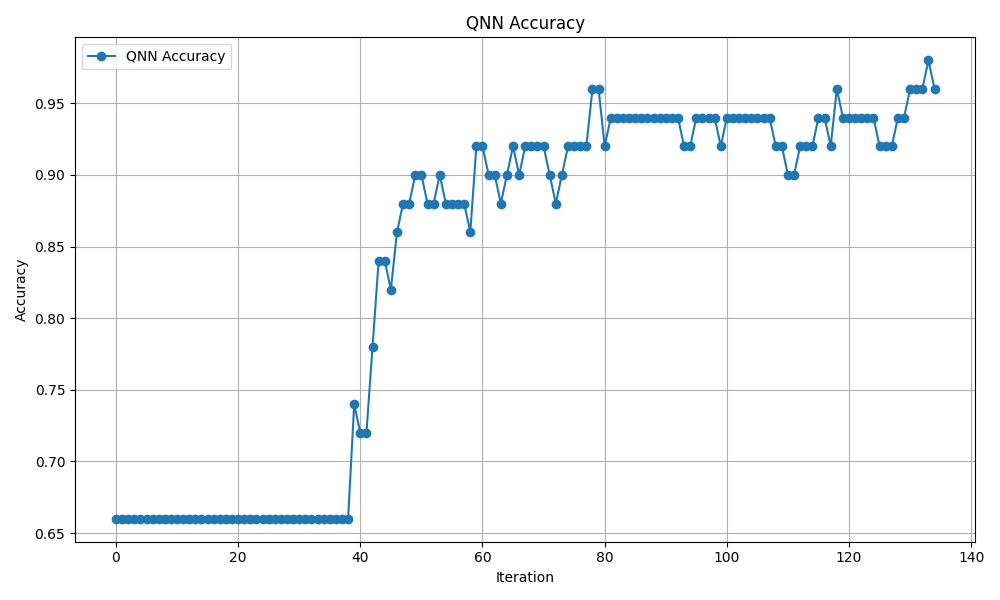}
        \label{fig:QNetAccuracyWisconsin}
    }
    \subfigure[Wisconsin Breast Cancer Accuracy Comparison]{%
        \includegraphics[width=0.45\textwidth]{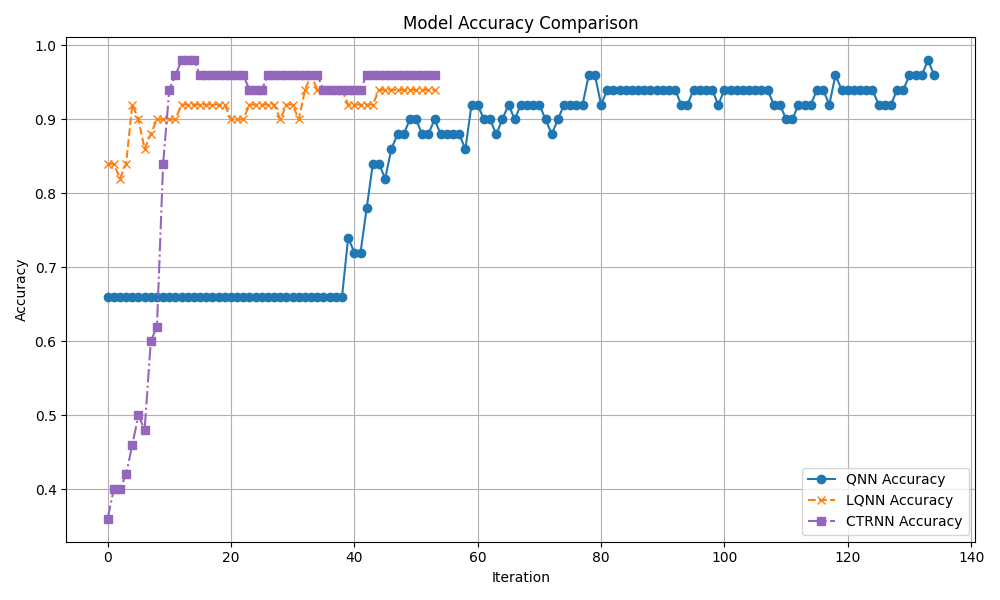}
        \label{fig:AccuracyComparisonWisconsin}
    }
    \caption{Accuracy Curves and Comparisons for Quantum Networks on Wisconsin Breast Cancer Dataset}
\end{figure}
According to Figure 9, all three models reached 95\% accuracy. However, the LQNet and CTRQNet took a significantly shorter time to converge, taking only 30 and 10 steps respectively. On the other hand the QNN only converges after 80 steps.
\begin{table}[H]
    \centering
    \resizebox{\textwidth}{!}{
        \begin{tabular}{|l|c|c|c|c|c|c|}
            \hline
            \textbf{Model} & \textbf{Accuracy (\%)} & \textbf{F1 Score} & \textbf{Precision Class 1 (\%)} & \textbf{Precision Class 0 (\%)} & \textbf{Recall Class 1 (\%)} & \textbf{Recall Class 0 (\%)} \\
            \hline
            CTRQNet & 98.25 & 98.25 & 99 & 98 & 99 & 98 \\
            QLNet   & 97.37 & 97.36 & 97 & 98 & 99 & 95 \\
            QNN    & 96.49 & 96.51 & 99 & 93 & 96 & 98 \\
            \hline
        \end{tabular}
    }
    \caption{Performance Metrics of Different Models on Wisconsin Breast Cancer Dataset}
    \label{tab:performance_metrics_wisconsin}
\end{table}
According to Table 3, the CTRQNet again outperforms the LQNet and the QNN by a slight margin.
\subsubsection{CIFAR 10}
The CIFAR 10 dataset consists of 200 training images, 100 validation images and 2000 testing images. We use the first two classes of the original dataset: Airplanes and Automobiles. For this dataset, we do not use preprocessing from pre-trained models.
\begin{figure}[H]
    \centering
    \subfigure[LQNet Loss Curve]{%
        \includegraphics[width=0.45\textwidth]{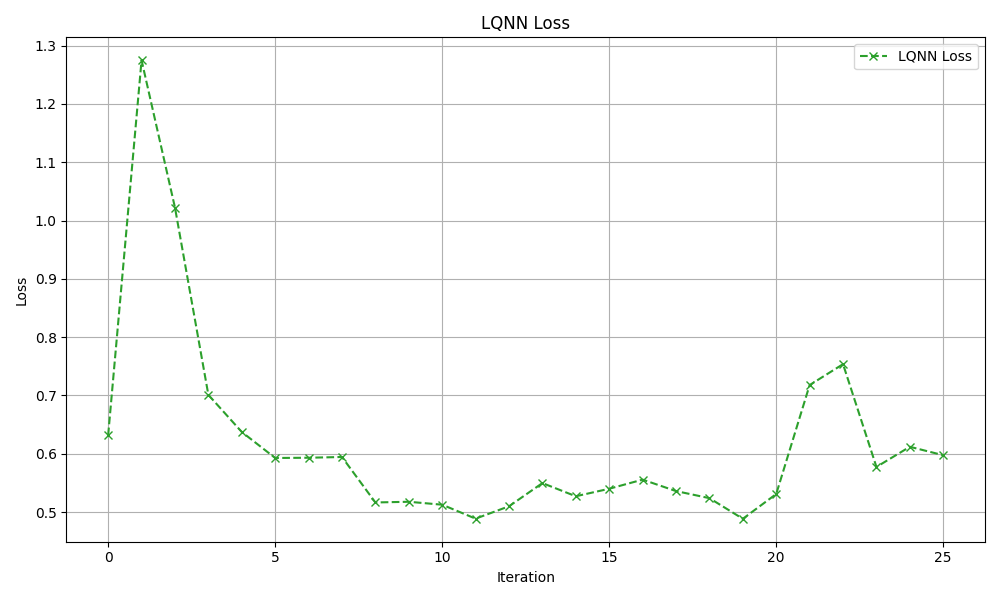}
        \label{fig:LQNetLossCIFAR}
    }
    \subfigure[CTRQNet Loss Curve]{%
        \includegraphics[width=0.45\textwidth]{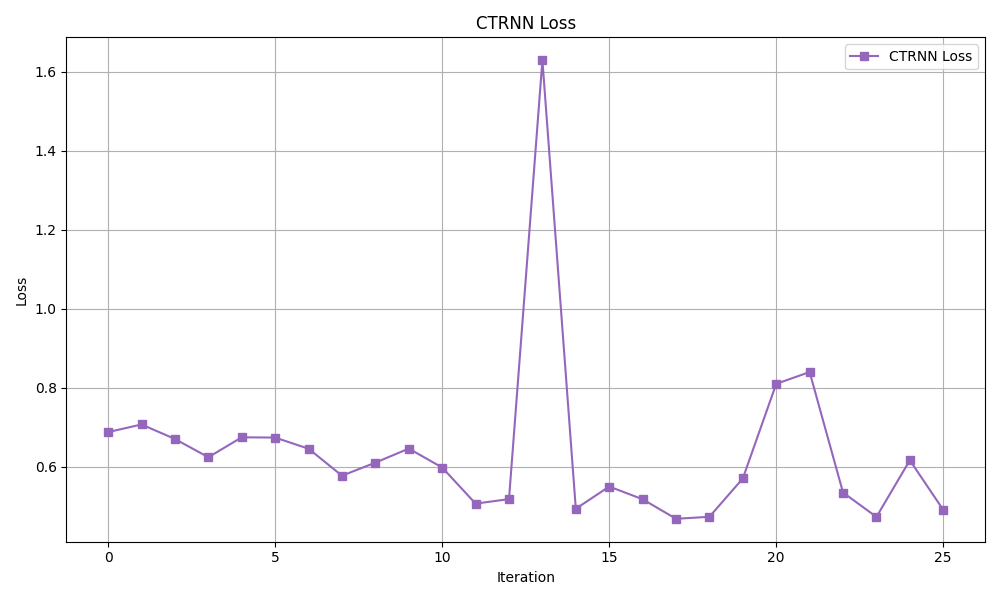}
        \label{fig:CTRQNetLossCIFAR}
    }\\
    \subfigure[QNN Loss Curve]{%
        \includegraphics[width=0.45\textwidth]{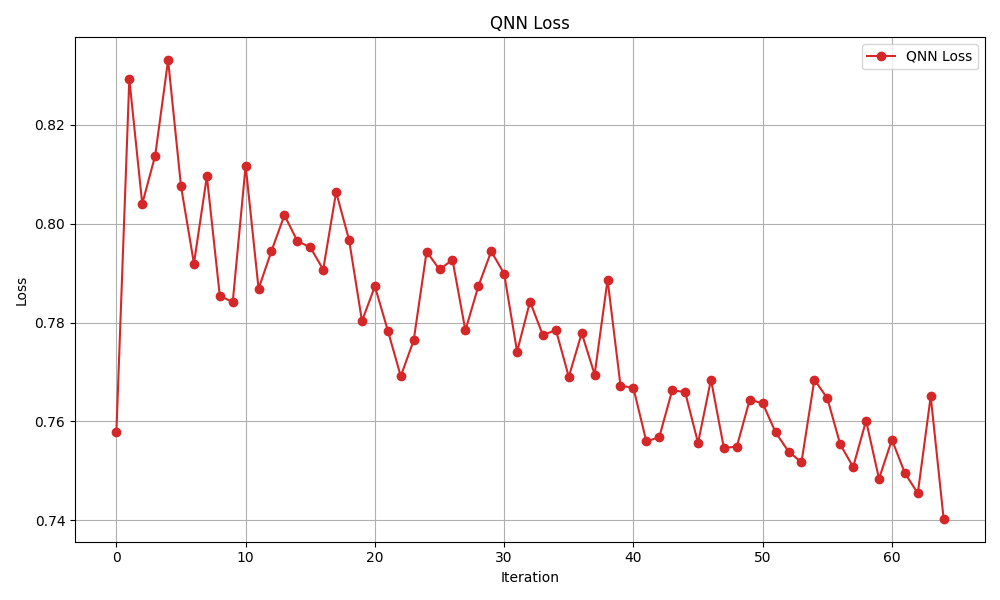}
        \label{fig:QNetLossCIFAR}
    }
    \subfigure[CIFAR 10 Loss Comparison]{%
        \includegraphics[width=0.45\textwidth]{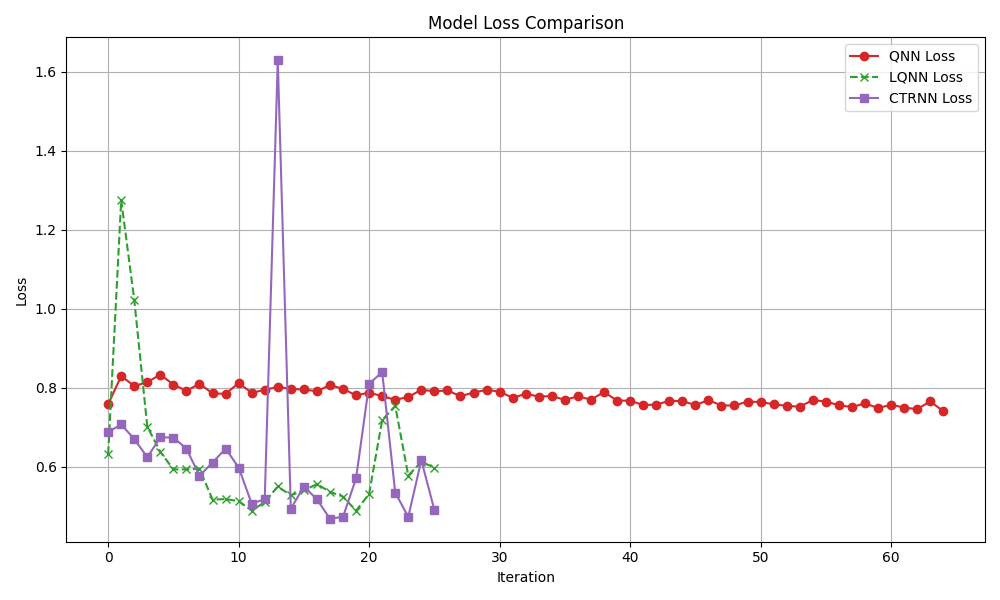}
        \label{fig:LossComparisonCIFAR}
    }
    \caption{Loss Curves and Comparisons for Quantum Networks on CIFAR 10}
\end{figure}
In Figure 10, the LQNet and CTRQNet demonstrate noticeable volatility in their loss curves; neither model can maintain optimal performance for long periods. The LQNet suffers a massive spike in its loss after two steps, and although it returns to its previous state and optimizes properly shortly after, it experiences another spike in loss after 23 steps. The CTRQNet optimizes initially but displays a massive upward spike after 13 steps before returning to an optimal state immediately afterward. We attribute the high volatility in both models to the difficulty of the task at hand. Meanwhile, the QNN cannot learn relationships underlying the data, which is reflected in its loss curve. After 65 steps, the QNN sees no noticeable difference compared to the initial loss. This task is too difficult for the QNN to learn anything, while the LQNet and CTRQNet are both able to learn meaningful relationships.

\begin{figure}[H]
    \centering
    \subfigure[LQNet Accuracy Curve]{%
        \includegraphics[width=0.45\textwidth]{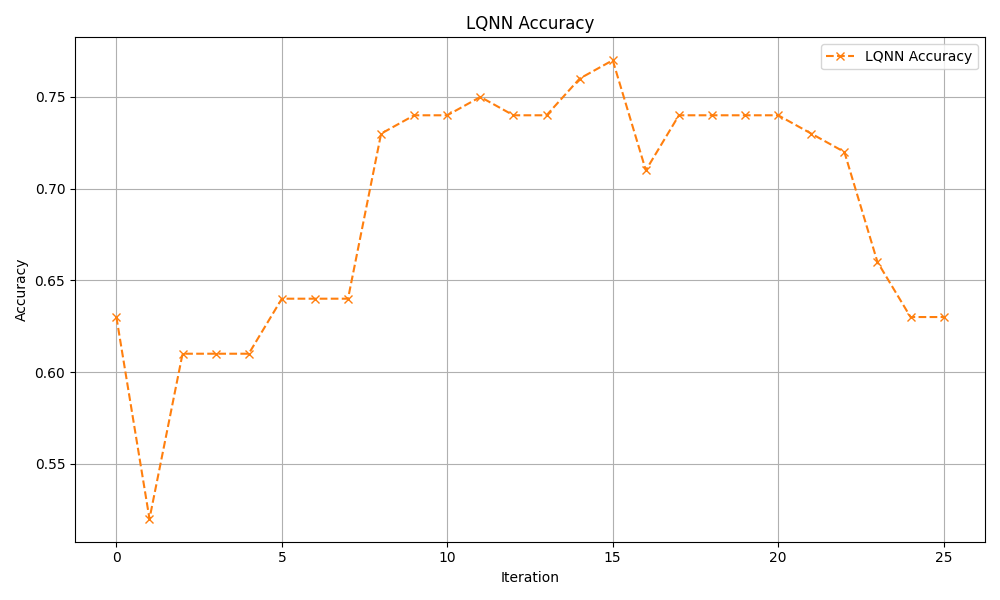}
        \label{fig:LQNetAccuracyCIFAR}
    }
    \subfigure[CTRQNet Accuracy Curve]{%
        \includegraphics[width=0.45\textwidth]{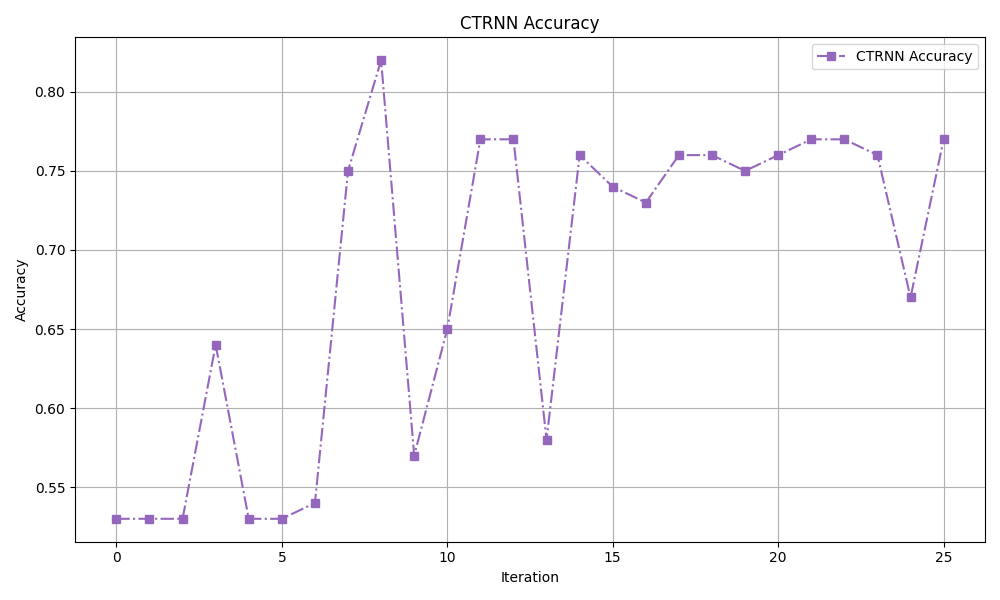}
        \label{fig:CTRQNetAccuracyCIFAR}
    }\\
    \subfigure[QNN Accuracy Curve]{%
        \includegraphics[width=0.45\textwidth]{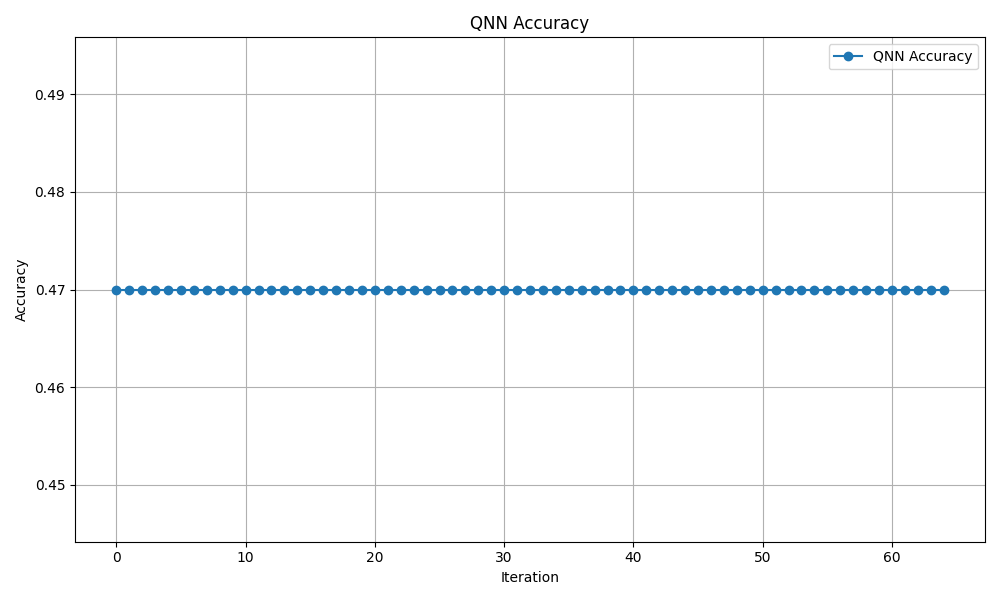}
        \label{fig:QNetAccuracyCIFAR}
    }
    \subfigure[CIFAR 10 Accuracy Comparison]{%
        \includegraphics[width=0.45\textwidth]{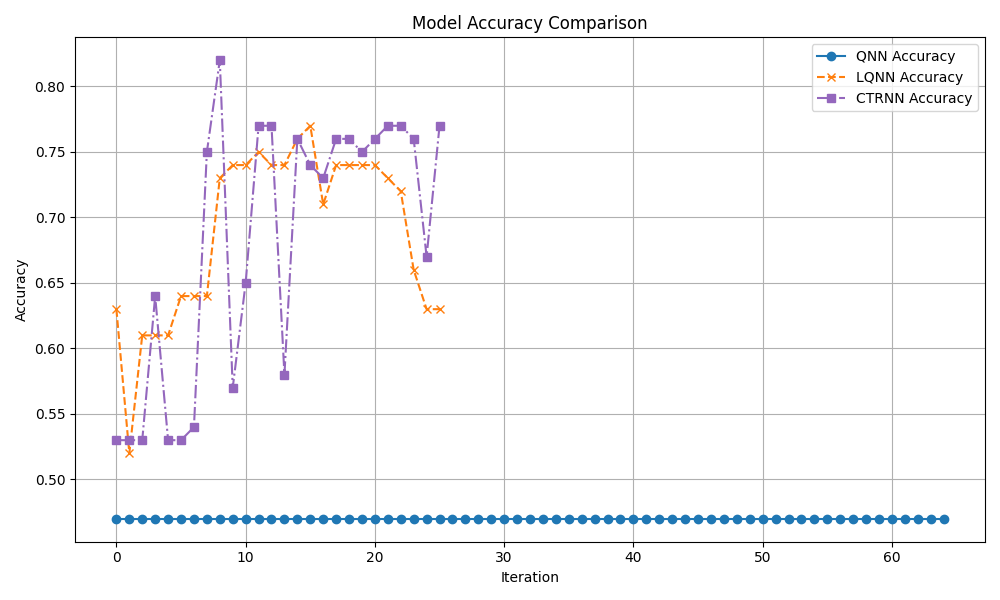}
        \label{fig:AccuracyComparisonCIFAR}
    }
    \caption{Accuracy Curves and Comparisons for Quantum Networks on CIFAR 10}
\end{figure}
In Figure 11, both the LQNet and CTRQNet show more volatility compared on this dataset in comparison to other baseline tests shown above. While both models are able to converge to an optimal state with satisfactory performance, they are unable to stay in the state consistently. We attribute this to overtraining as we use a classical set of hyperparameters to train these models. Both models achieve a high of 75-80\% accuracy on the validation set. The QNN is unable to learn anything meaningful and maintains a constant accuracy of 47\%. Our new models are able to learn tasks that prove to be too difficult for QNNs. 

\begin{table}[H]
    \centering
    \resizebox{\textwidth}{!}{
        \begin{tabular}{lcccccc}
            \hline
            \textbf{Model} & \textbf{Accuracy} & \textbf{F1 Score} & \textbf{Precision Class 1} & \textbf{Precision Class 0} & \textbf{Recall Class 1} & \textbf{Recall Class 0} \\
            \hline
            CTRQNet & 75.8 & 75.47 & 81 & 74 & 64 & 87 \\
            QLNet & 70.35 & 68.14 & 63 & 93 & 97 & 60 \\
            QNN & 50 & 33.3 & 50 & 0 & 100 & 0 \\
            \hline
        \end{tabular}
    }
    \caption{Performance Metrics for Quantum Networks on CIFAR-10}
    \label{tab:performance_metrics_CIFAR}
\end{table}
Table 4 shows that the CTRQNet has the best performance across all metrics on the test dataset, followed closely by the LQNet, achieving 75.8\% and 70.35\% accuracy, respectively. In comparison, the QNN is unable to learn patterns underlying the data, reflected in the QNN's accuracy of 50\% due to the QNN outputting the same class across all test samples. We attribute this to the QNN having a rigid hidden state, leading it to learn only the latent representation of the data rather than the patterns underlying it.

The CTRQNet and LQNet can learn the CIFAR 10 dataset and converge rapidly to a satisfactory optimal state. Due to the large samples in the CIFAR 10 dataset, all three models learn a latent space representation of the data rather than the direct patterns underlying the data. While the CTRQNet and LQNet achieve satisfactory results when tested, these results can be further improved by letting the models learn the patterns behind the data rather than a latent representation of it. Additionally, we scale the number of parameters in these models since our models, the LQNet and CTRQNet, only utilize 224,000 parameters. In contrast, convolutional neural networks, such as Resnet 151 \cite{Resnet}, utilize over 60 million parameters, a large fraction of which are dedicated to learning only the latent representation of the data. Due to the huge amount of parameters these models possess, there are enough parameters to learn the patterns behind the data and the latent representation. The majority of the parameters in our models are dedicated to learning the latent representation of the data, leaving an insufficient amount to learn the patterns behind the data due to the small number of parameters. In conclusion, We are unable to scale our models due to intensive computational complexity within pseudo-quantum computing.
\subsubsection{CIFAR 10 Feature Extraction}
Due to the computational complexity required to model the CIFAR 10 dataset accurately, we use a pre-trained model of residual networks to preprocess the images. The preprocessing model in question is Resnet 151. With the addition of this model, we can help the models learn deeper patterns due to Resnet taking care of computational complexity and leaving the CTRQNet and LQNet to learn the patterns underlying the data rather than learning a latent space representation of the data.

Based on Figure 12, both the CTRQNet and LQNet display a similar training curve with relatively low volatility. Additionally, both models converge rapidly to an optimal state and maintain their performance after reaching this state. However, they only maintain this state for 30 steps after reaching the optimal state before experiencing a slow increase in loss. This is attributed to the usage of a classical set of hyperparameters on a quantum model. The QNN has a prolonged learning process, showing no real progress in its ability to learn the task. Our models are able to perform well on difficult datasets, which prove to be too difficult for QNNs to perform on.
\begin{figure}[H]
    \centering
    \subfigure[LQNet Loss Curve]{%
        \includegraphics[width=0.38\textwidth]{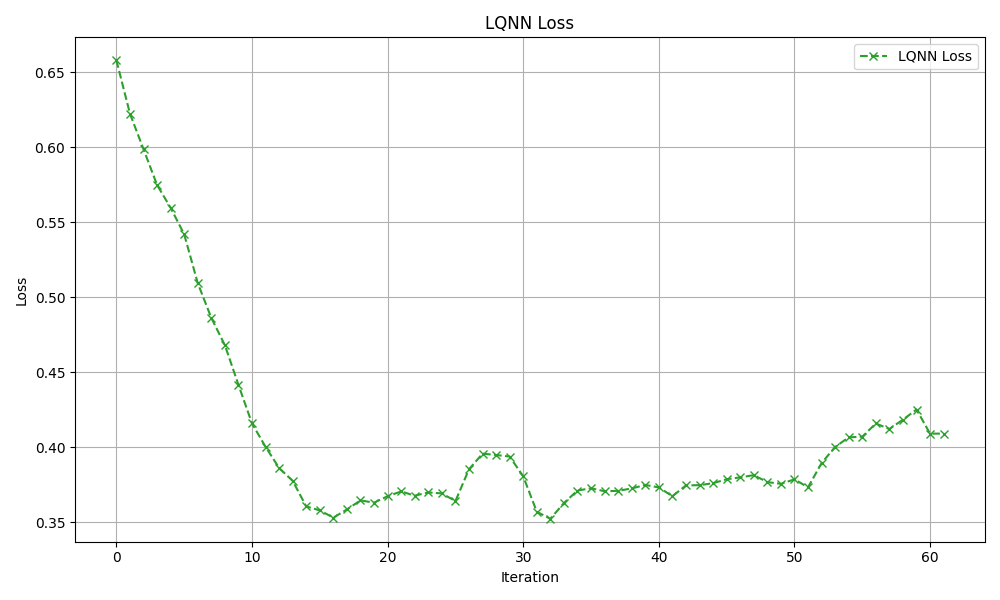}
        \label{fig:LQNetLossCIFAR10Feature}
    }
    \subfigure[CTRQNet Loss Curve]{%
        \includegraphics[width=0.38\textwidth]{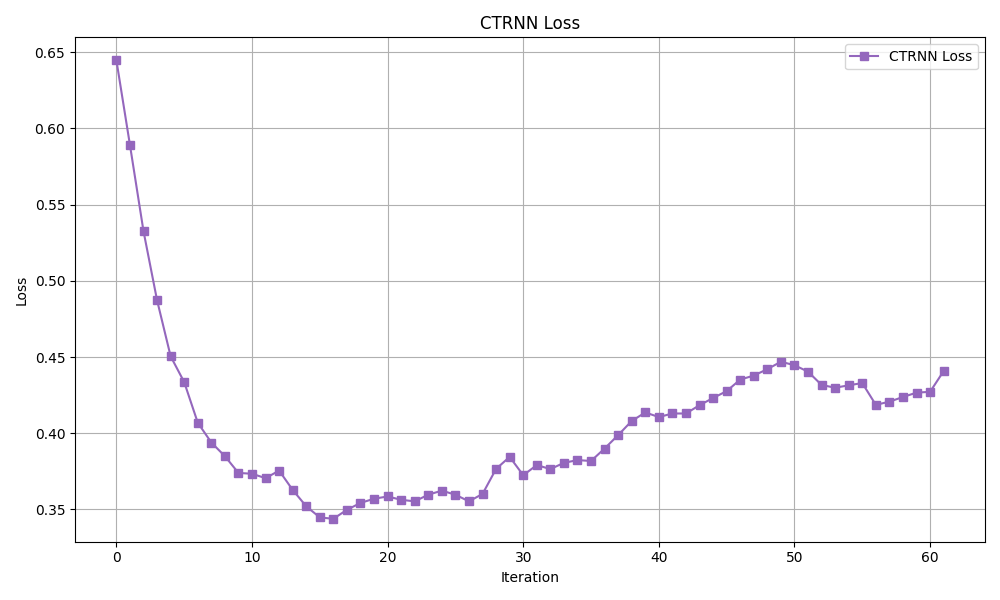}
        \label{fig:CTRQNetLossCIFAR10}
    }
    \subfigure[QNN Loss Curve]{%
        \includegraphics[width=0.38\textwidth]{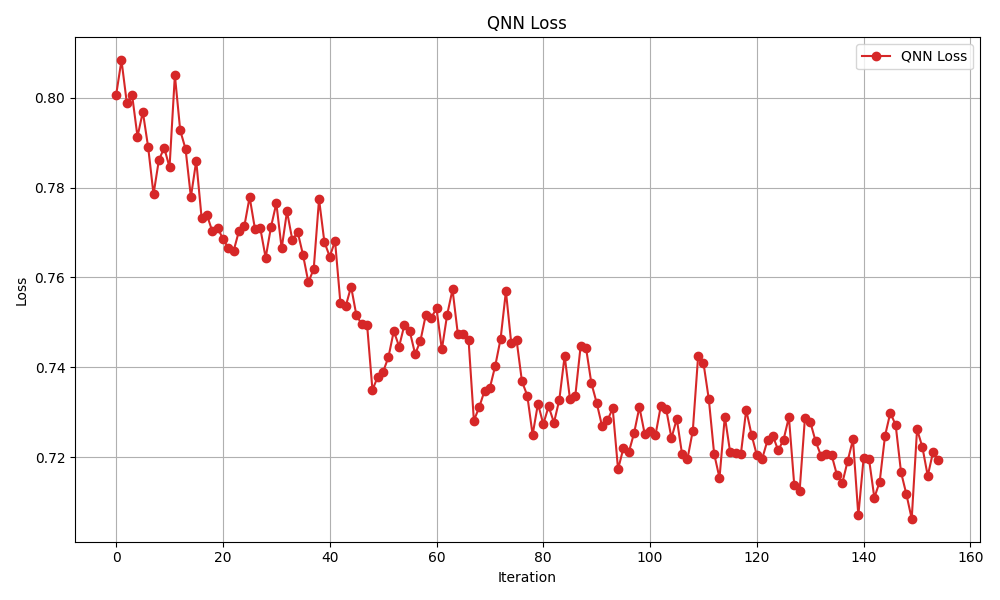}
        \label{fig:QNetLossCIFAR10Feature}
    }
    \subfigure[CIFAR 10 with Downsampling Layers Loss Comparison]{%
        \includegraphics[width=0.38\textwidth]{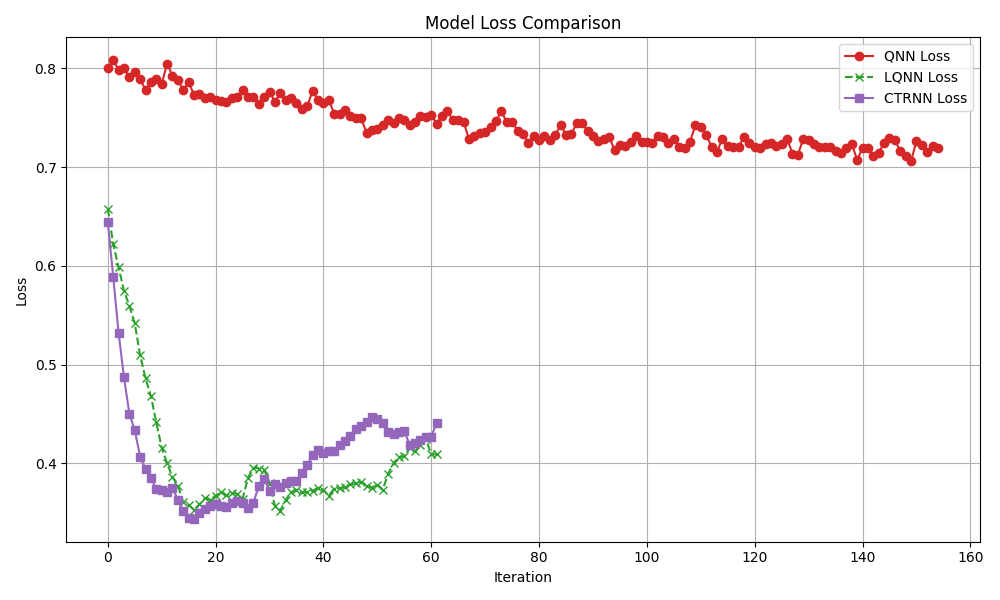}
        \label{fig:LossComparisonCIFAR10Feature}
    }
    \caption{Loss Curves and Comparisons for Quantum Networks on CIFAR 10 with Downsampling Layers}
\end{figure}

\begin{figure}[H]
    \centering
    \subfigure[LQNet Accuracy Curve]{%
        \includegraphics[width=0.38\textwidth]{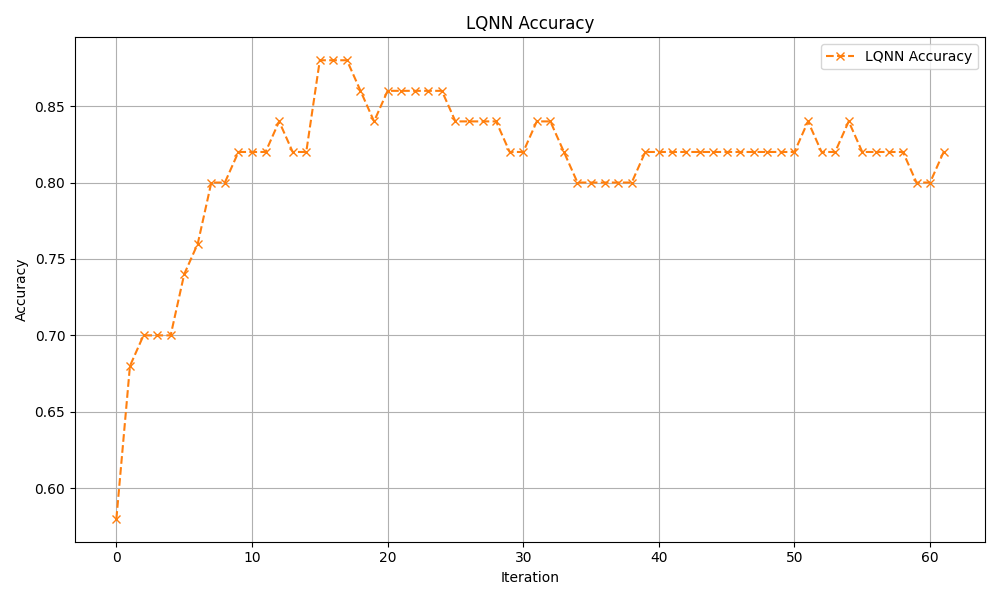}
        \label{fig:LQNetAccuracyCIFAR10Featured}
    }
    \subfigure[CTRQNet Accuracy Curve]{%
        \includegraphics[width=0.38\textwidth]{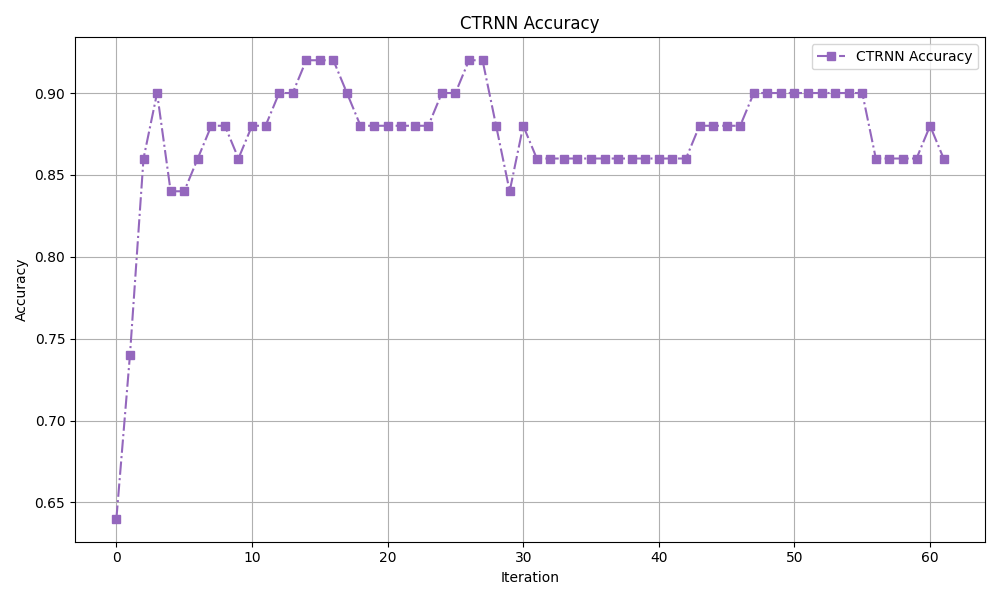}
        \label{fig:CTRQNetAccuracyCIFAR10Featured}
    }\\
    \subfigure[QNN Accuracy Curve]{%
        \includegraphics[width=0.38\textwidth]{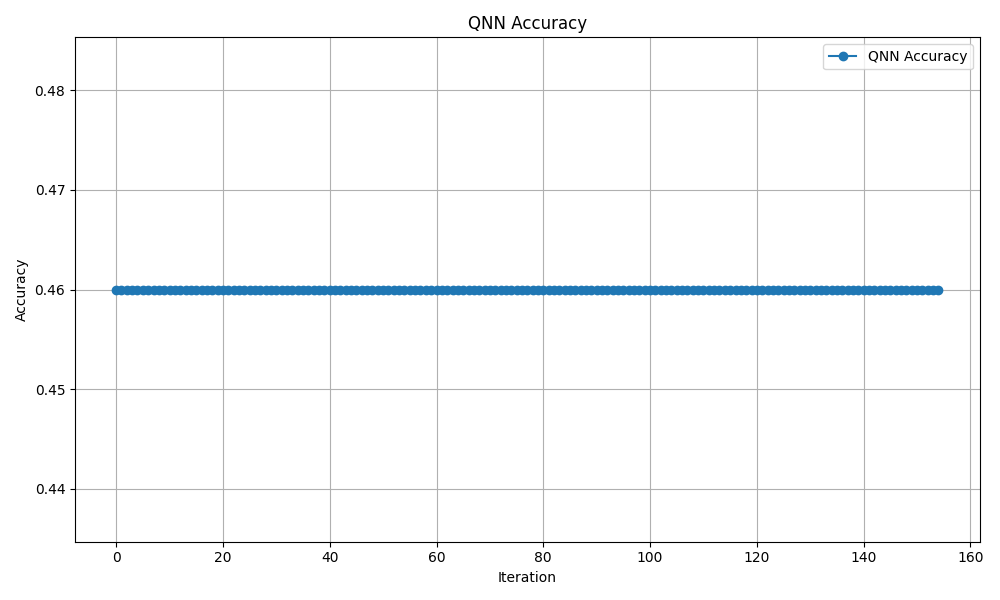}
        \label{fig:QNetAccuracyCIFAR10Featured}
    }
    \subfigure[CIFAR 10 Accuracy Comparison]{%
        \includegraphics[width=0.38\textwidth]{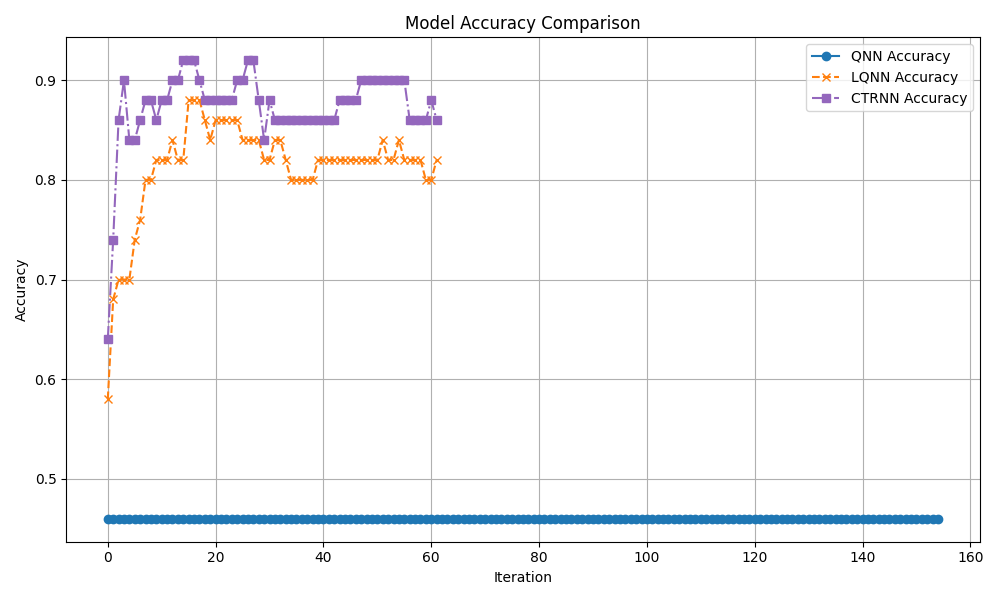}
        \label{fig:AccuracyComparisonCIFAR10Featured}
    }
    \caption{Accuracy Curves and Comparisons for Quantum Networks on CIFAR 10 Feature}
\end{figure}
Looking at Figure 13, the QNN is unable to learn meaningful information underlying the data within a reasonable training time, unlike the CTRQNet and LQNet, which are both able to converge to an optimal state quickly. Both models converge at around 15 steps, maintain their performance after reaching this peak, and have lots of movement in terms of accuracy but maintain optimal performance, ensuring that the state of these models does not stay constant. This is a benefit as common issues that arise in classical and quantum machine learning, such as the vanishing gradient, will not affect our models.
Below, we present the ROC curves and confusion matrices.

\begin{figure}[H]
    \centering
    \subfigure[LQNet ROC Curve]{%
        \includegraphics[width=0.3\textwidth]{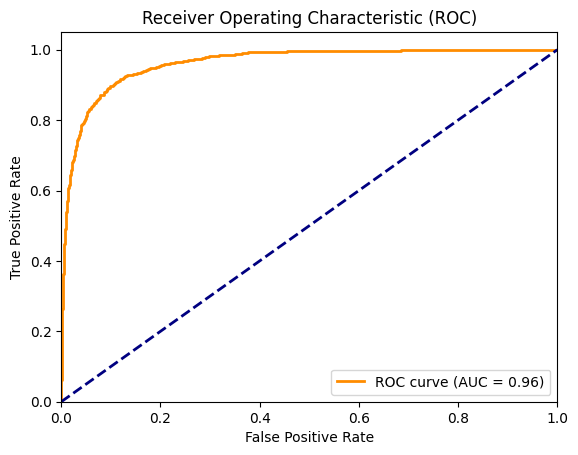}
        \label{fig:Image1}
    }
    \subfigure[CTRQNet ROC Curve]{%
        \includegraphics[width=0.3\textwidth]{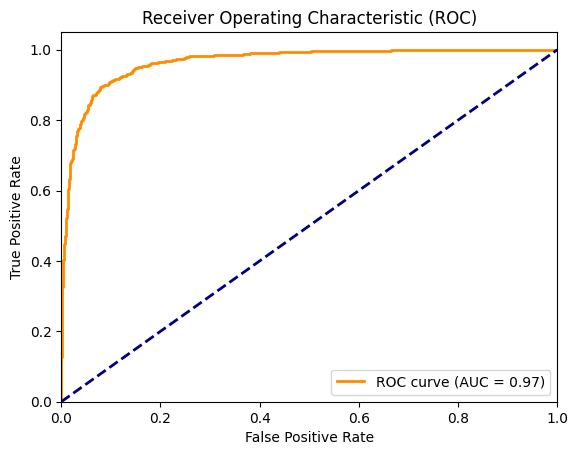}
        \label{fig:Image2}
    }
    \subfigure[QNN ROC Curve]{%
        \includegraphics[width=0.3\textwidth]{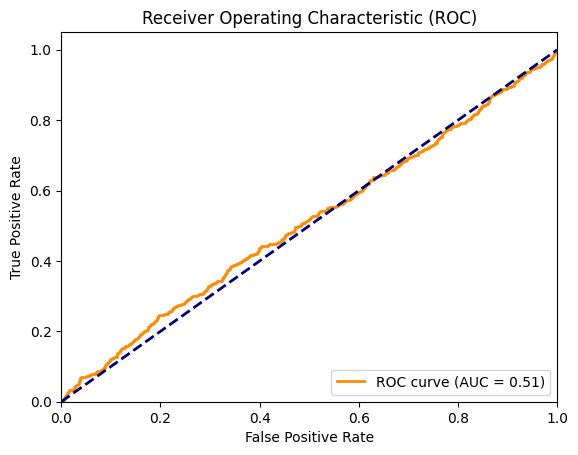}
        \label{fig:Image3}
    }\\
    \subfigure[LQNet Confusion Matrix]{%
        \includegraphics[width=0.3\textwidth]{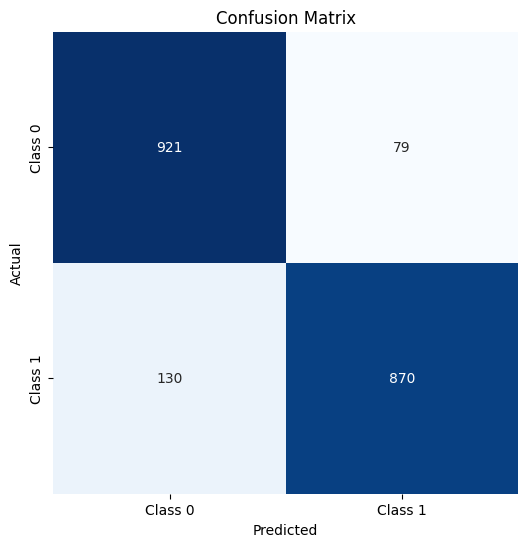}
        \label{fig:Image4}
    }
    \subfigure[CTRQNet Confusion Matrix]{%
        \includegraphics[width=0.3\textwidth]{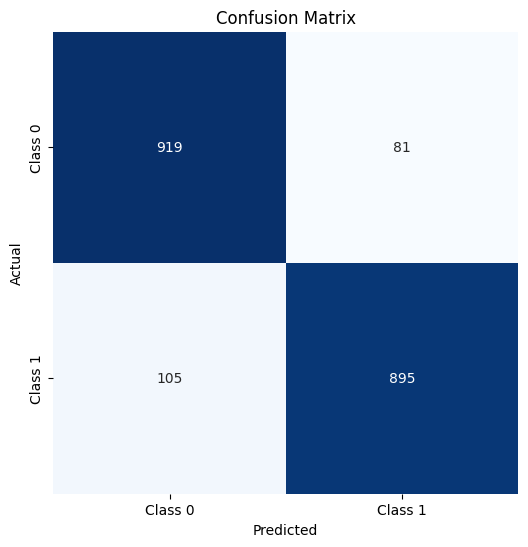}
        \label{fig:Image5}
    }
    \subfigure[QNN Confusion Matrix]{%
        \includegraphics[width=0.3\textwidth]{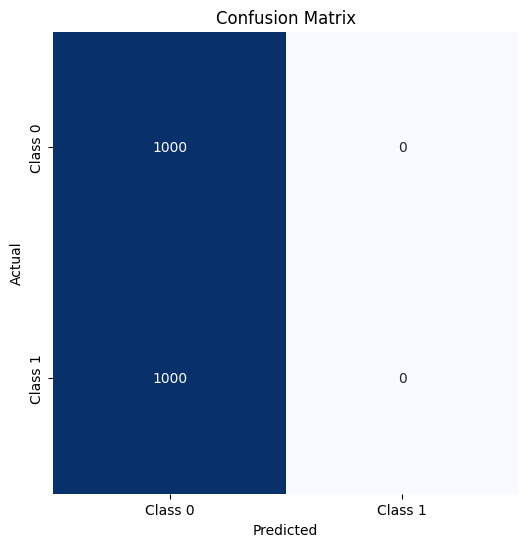}
        \label{fig:Image6}
    }
    \caption{ROC Curve and Confusion Matrix Comparisons on CIFAR 10 Feature}
\end{figure}

\begin{table}[H]
    \centering
    \resizebox{\textwidth}{!}{
        \begin{tabular}{|l|c|c|c|c|c|c|}
            \hline
            \textbf{Model} & \textbf{Accuracy (\%)} & \textbf{F1 Score} & \textbf{Precision Class 1 (\%)} & \textbf{Precision Class 0 (\%)} & \textbf{Recall Class 1 (\%)} & \textbf{Recall Class 0 (\%)} \\
            \hline
            CTRQNet & 90.25 & 90.25 & 91 & 89 & 92 & 89 \\
            QLNet   & 90.60 & 90.60 & 92 & 89 & 89 & 92 \\
            QNN    & 50.00 & 33.00 & 0  & 50 & 100 & 0  \\
            \hline
        \end{tabular}
    }
    \caption{Performance Metrics for Quantum Networks on CIFAR 10 with Downsampling Layers}
    \label{tab:performance_metrics_CIFAR10Feature}
\end{table}
According to Table 5, the CTRQNet and QLNet have performed similarly, reaching 90\% accuracy, whereas the QNN has an accuracy of 50\%. Once again, the QNN outputted the same class across all test samples. Even after preprocessing the dataset, CIFAR 10 remains too complex for the QNN to learn.

After the CIFAR 10 dataset was preprocessed, the CTRQNet and LQNet showed a considerable improvement in accuracy, going from a 73.1\% average to a 90\% average. On the other hand, the QNN showed no such improvements, maintaining the same 50\% accuracy as when they used the original CIFAR 10 dataset. While having a downward slope, its loss graph remains volatile, showing that it has trouble learning the CIFAR 10 dataset even after preprocessing.

\section{Conclusion and Future Research}
\subsection{New Findings}
In this research, we develop two new state-of-the-art quantum machine learning models, LQNets and CTRQNets, which leverage the continuous dynamics from liquid networks and continuous-time recurrent networks in order to make the structure of the quantum network more dynamic and widen its ability to learn complex tasks. We tested these new models on multiple datasets, and our results show that the models outperform QNNs on all five datasets. Using the MNIST dataset, the models also outperformed the results found in \cite{QRESNET}. In both CIFAR 10 datasets, the LQNet and CTRQNet achieved satisfactory results.

\subsection{Future Research}
Throughout the testing of these models, we observe that our newly created quantum models are extremely prone to changes in performance at intermediate points throughout the training. However, this does not inhibit the models' ability to converge. As this research was unable to determine the extent of the relationship between the performance and its hyperparameters, future research might consider how the networks perform under different hyperparameter configurations. Due to the limited number of qubits available, we are unable to test these models using larger quantum circuits. With access to quantum hardware, we would be able to modify our proposed models, CTRQNets and LQNets.
\bibliographystyle{IEEEtran}
\bibliography{Sources.bib}
\section{Appendix}
Here we include additional ROC curves and confusion matrices for our proposed models. 
\begin{figure}[H]
    \centering
    \subfigure[LQNet ROC Curve]{%
        \includegraphics[width=0.44\textwidth]{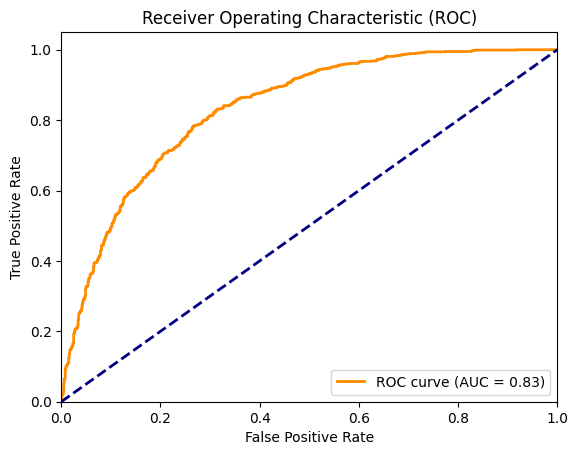}
        \label{fig:Image7}
    }
    \subfigure[LQNet Confusion Matrix]{%
        \includegraphics[width=0.33\textwidth]{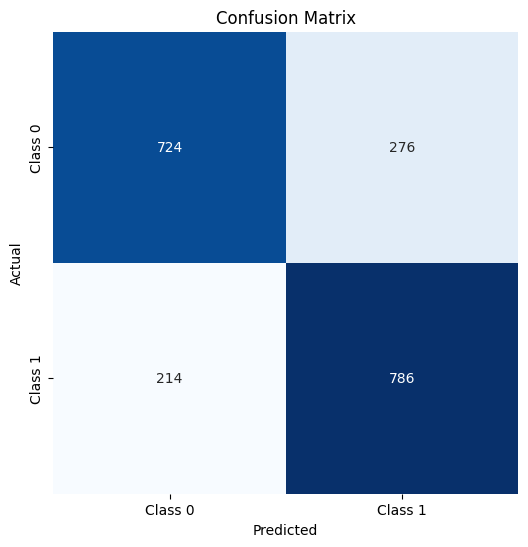}
        \label{fig:Image8}
    }\\
    \subfigure[CTRQNet ROC Curve]{%
        \includegraphics[width=0.44\textwidth]{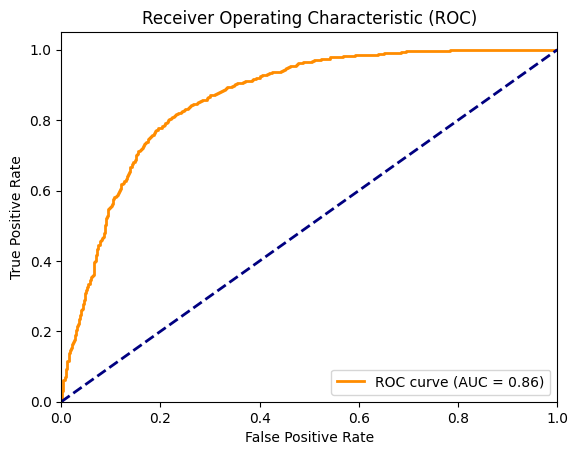}
        \label{fig:Image9}
    }
     \subfigure[CTRQNet Confusion Matrix]{%
        \includegraphics[width=0.33\textwidth]{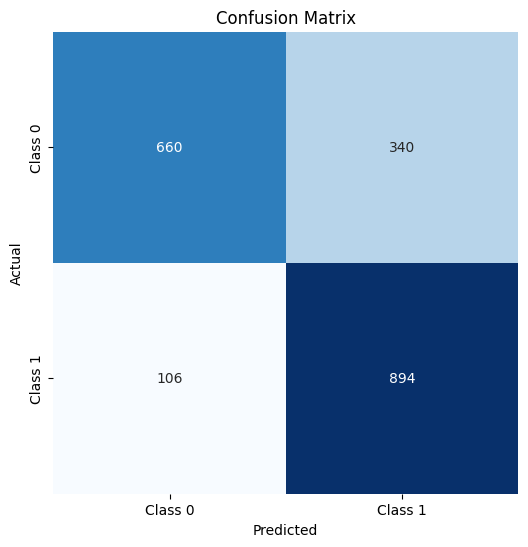}
        \label{fig:Image10}
    }\\
    \subfigure[QNN ROC Curve]{%
        \includegraphics[width=0.44\textwidth]{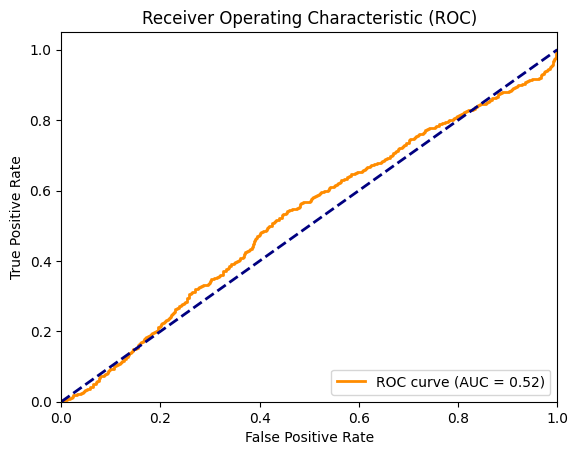}
        \label{fig:Image11}
    } 
    \subfigure[QNN Confusion Matrix]{%
        \includegraphics[width=0.33\textwidth]{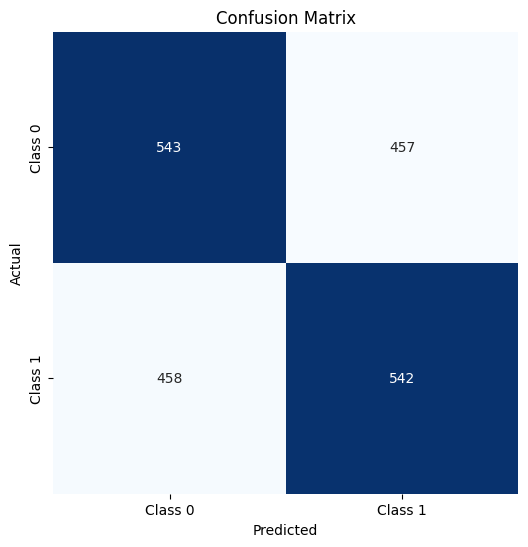}
        \label{fig:Image12}
    }
    \caption{ROC Curve and Confusion Matrix Comparisons on CIFAR 10}
\end{figure}

\begin{figure}[H]
    \centering
    \subfigure[LQNet ROC Curve]{%
        \includegraphics[width=0.44\textwidth]{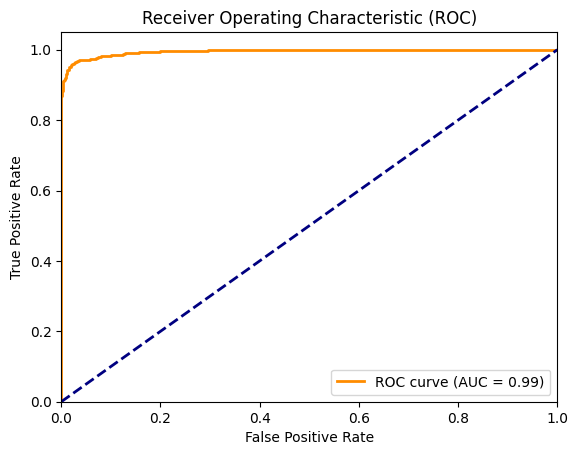}
        \label{fig:Image13}
    }
    \subfigure[LQNet Confusion Matrix]{%
        \includegraphics[width=0.33\textwidth]{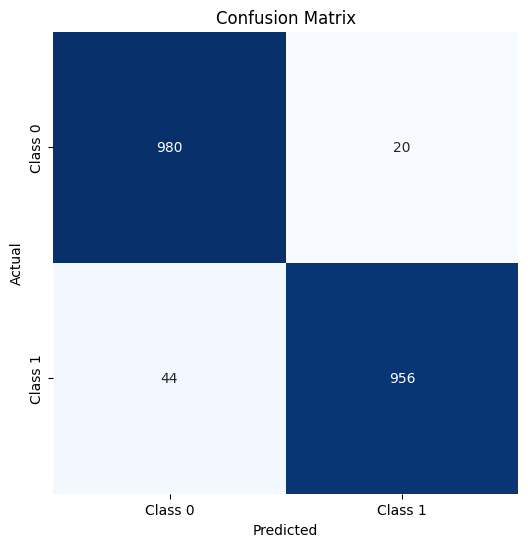}
        \label{fig:Image14}
    }\\
    \subfigure[CTRQNet ROC Curve]{%
        \includegraphics[width=0.44\textwidth]{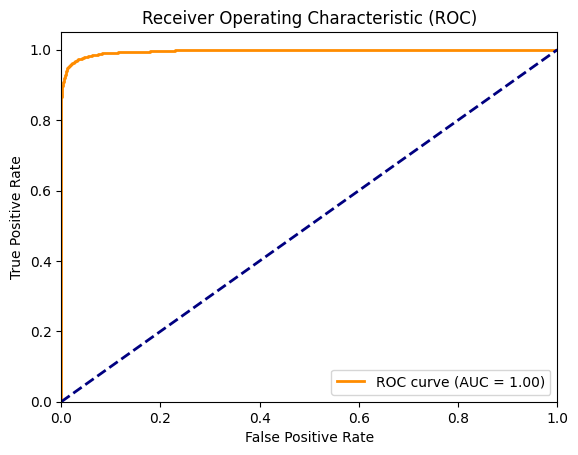}
        \label{fig:Image15}
    }
    \subfigure[CTRQNet Confusion Matrix]{%
        \includegraphics[width=0.33\textwidth]{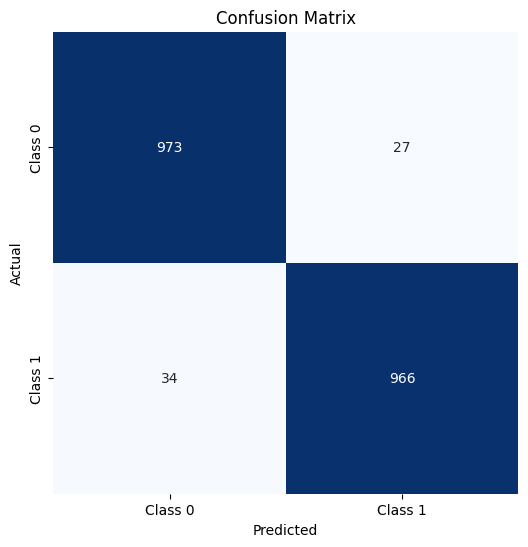}
        \label{fig:Image16}
    }\\
    \subfigure[QNN ROC Curve]{%
        \includegraphics[width=0.44\textwidth]{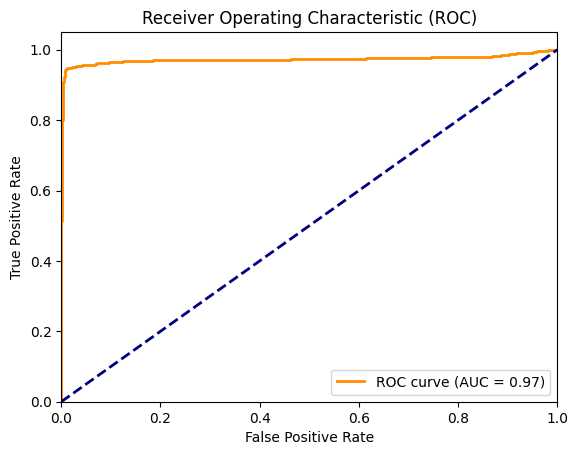}
        \label{fig:Image17}
    }    
    \subfigure[QNN Confusion Matrix]{%
        \includegraphics[width=0.33\textwidth]{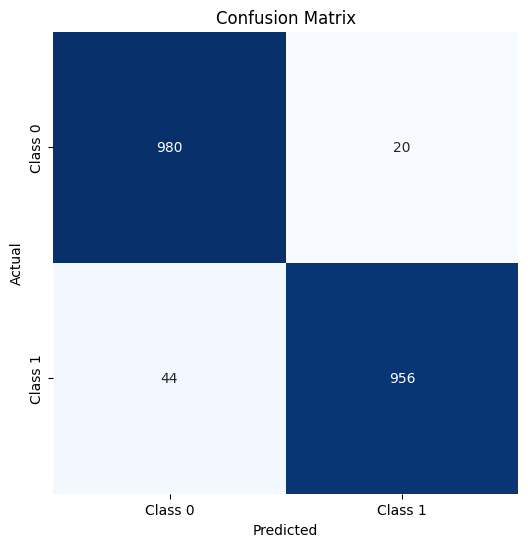}
        \label{fig:Image18}
    }
    \caption{ROC Curve and Confusion Matrix Comparisons on FMNIST}
\end{figure}

\begin{figure}[H]
    \centering
    \subfigure[LQNet ROC Curve]{%
        \includegraphics[width=0.44\textwidth]{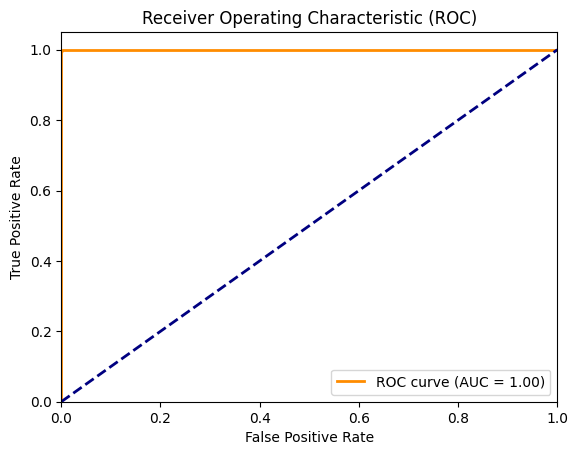}
        \label{fig:Image19}
    }
    \subfigure[LQNet Confusion Matrix]{%
        \includegraphics[width=0.33\textwidth]{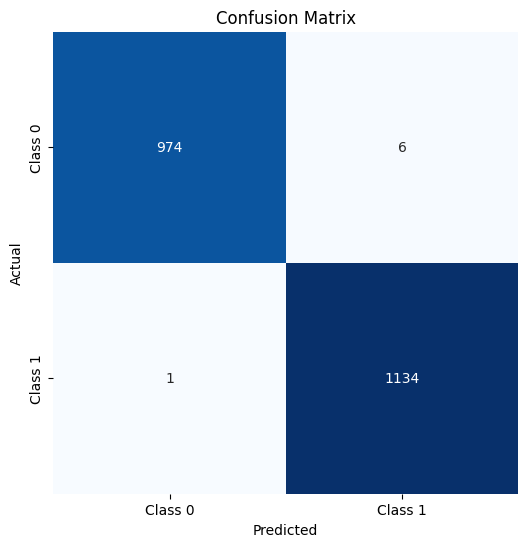}
        \label{fig:Image20}
    }\\
    \subfigure[CTRQNet ROC Curve]{%
        \includegraphics[width=0.44\textwidth]{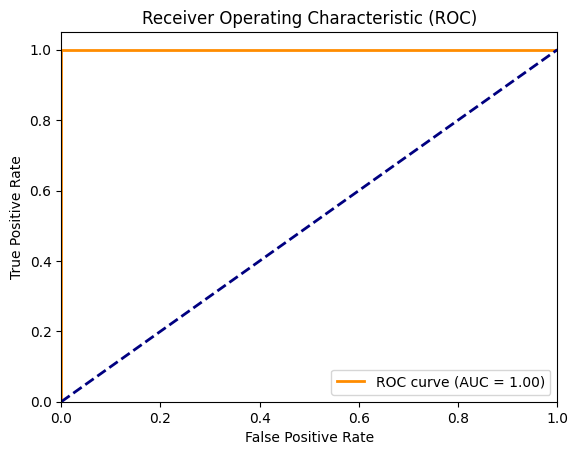}
        \label{fig:Image21}
    }
    \subfigure[CTRQNet Confusion Matrix]{%
        \includegraphics[width=0.33\textwidth]{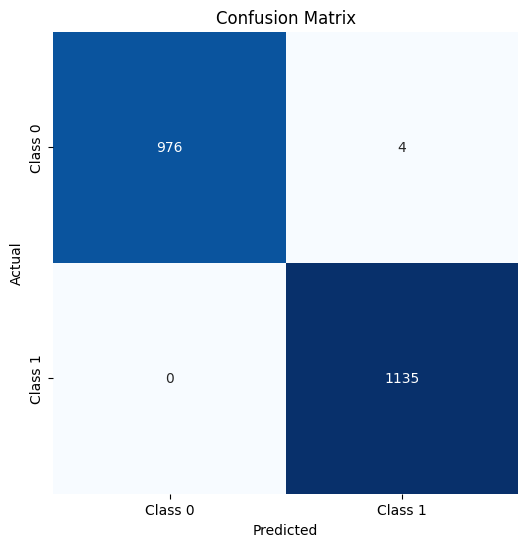}
        \label{fig:Image22}
    }\\
    \subfigure[QNN ROC Curve]{%
        \includegraphics[width=0.44\textwidth]{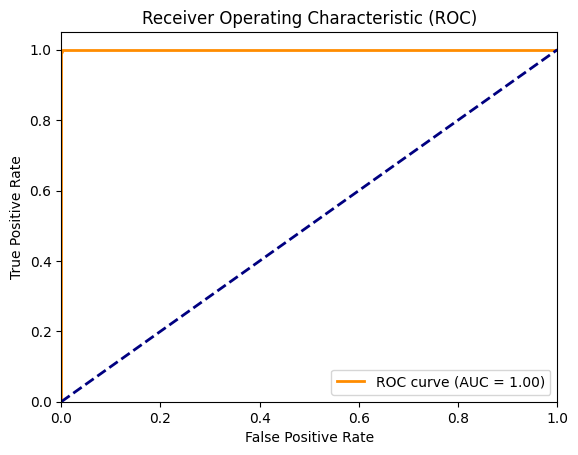}
        \label{fig:Image23}
    }
    \subfigure[QNN Confusion Matrix]{%
        \includegraphics[width=0.33\textwidth]{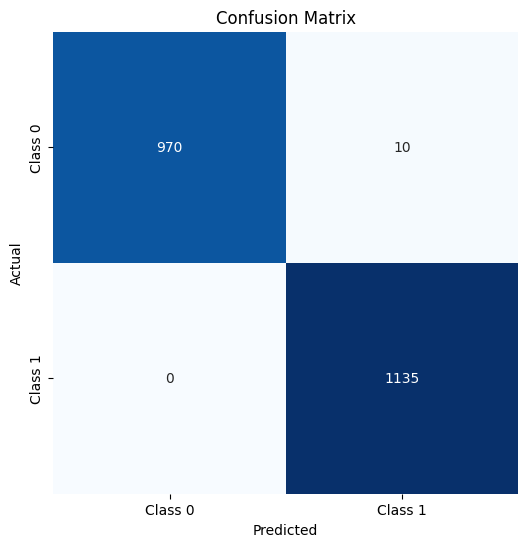}
        \label{fig:Image24}
    }
    \caption{ROC Curve and Confusion Matrix Comparisons on MNIST}
\end{figure}

\begin{figure}[H]
    \centering
    \subfigure[LQNet ROC Curve]{%
        \includegraphics[width=0.44\textwidth]{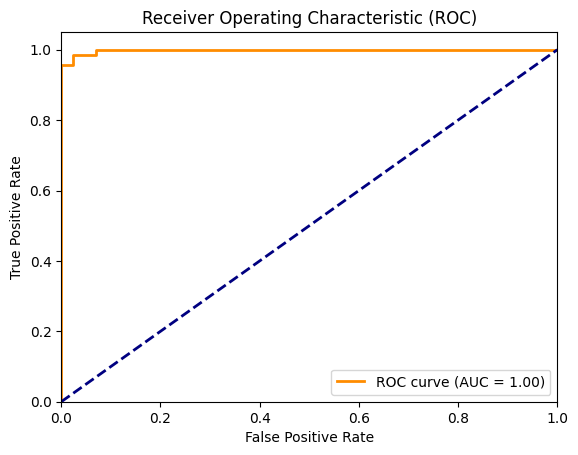}
        \label{fig:Image25}
    }
    \subfigure[LQNet Confusion Matrix]{%
        \includegraphics[width=0.33\textwidth]{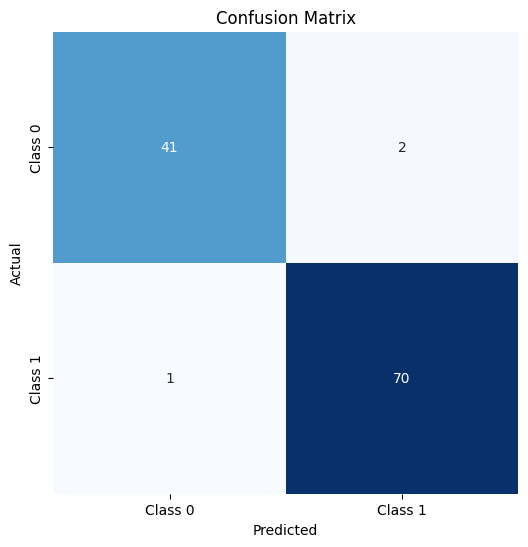}
        \label{fig:Image26}
    }\\
    \subfigure[CTRQNet ROC Curve]{%
        \includegraphics[width=0.44\textwidth]{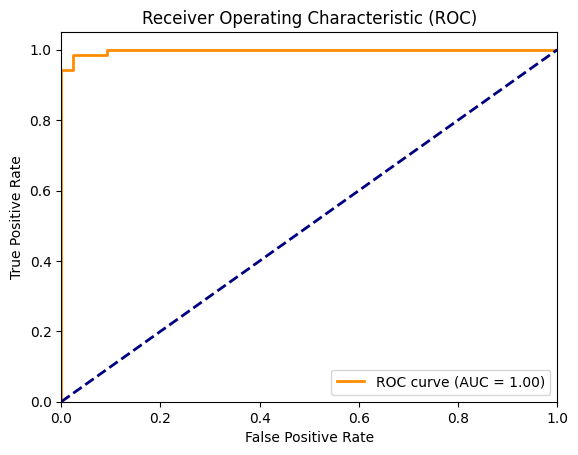}
        \label{fig:Image27}
    }
    \subfigure[CTRQNet Confusion Matrix]{%
        \includegraphics[width=0.33\textwidth]{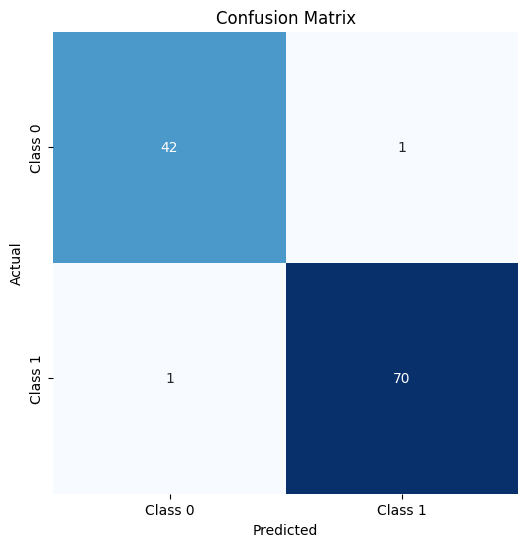}
        \label{fig:Image28}
    }\\
    \subfigure[QNN ROC Curve]{%
        \includegraphics[width=0.44\textwidth]{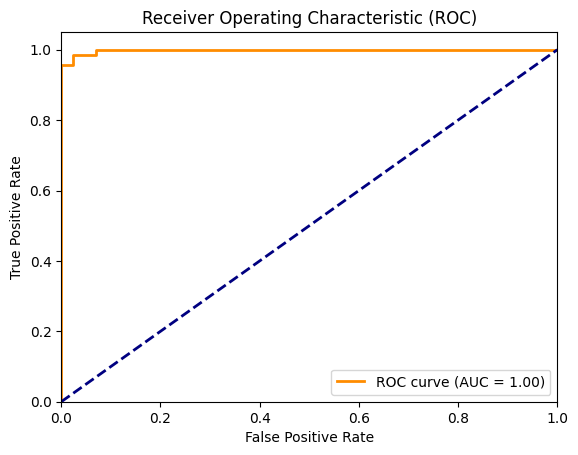}
        \label{fig:Image29}
    }
    \subfigure[QNN Confusion Matrix]{%
        \includegraphics[width=0.33\textwidth]{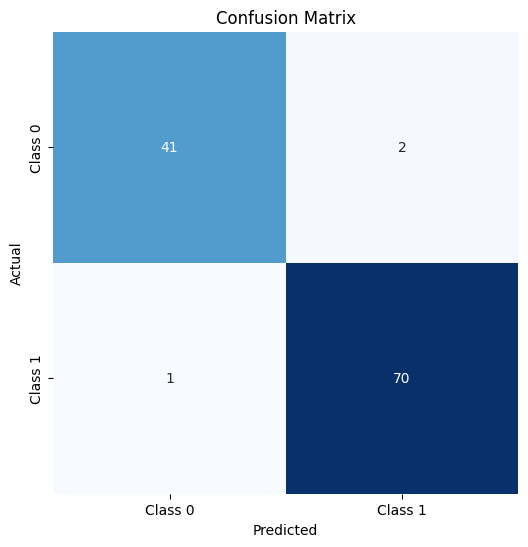}
        \label{fig:Image30}
    }
    \caption{ROC Curve and Confusion Matrix Comparisons on Wisconsin Breast Cancer Dataset}
\end{figure}

\end{document}